\documentclass[letterpaper, 10 pt, conference
]{ieeeconf}  

\IEEEoverridecommandlockouts                              

\usepackage{siunitx}

\usepackage{graphicx}

\usepackage{ascmac}
\usepackage{graphics} 
\usepackage{epsfig} 
\usepackage{amsfonts}
\usepackage{amsmath} 
\usepackage{amssymb}  
\usepackage{bm}
\usepackage{comment}
\usepackage{subfigure}
\usepackage{latexsym}
\usepackage{multicol}
\usepackage{float}
\usepackage{here}
\usepackage{url}
\usepackage{physics}

\title{\LARGE \bf
Application of Battery Storage to Switching Predictive Control of Power Distribution Systems Including Road Heating*
}

\author{Chiaki Kojima$^{1}$, Yuya Muto$^{2}$, Hikaru Akutsu$^{1}$, Rinnosuke Shima$^{1}$, and Yoshihiko Susuki$^{3}$
\thanks{*This work was supported by Azbil Yamatake General Foundation and JSPS KAKENHI Grant Numbers 20K04552, 23K03910.}
\thanks{$^{1}$Chiaki Kojima, Hikaru Akutsu, and Rinnosuke Shima are with Department of Electrical and 
Electronic Engineering, Faculty of Engineering, Toyama Prefectural 
University, 5180, Kurokawa, Imizu, Toyama, 939-0398, Japan
        {\tt\small chiaki@pu-toyama.ac.jp}}%
\thanks{$^{2}$Yuya Muto is with Department of Electrical and Computer Engineering, Graduate School of Engineering, Toyama Prefectural University, 5180, Kurokawa, Imizu, Toyama, 939-0398, Japan
        }%
\thanks{$^{3}$Department of Electrical Engineering, 
Graduate School of Engineering, Kyoto-University, Kyoto-Daigaku-Katsura, Nishikyo-ku, Kyoto 615-8510, Japan
        }%
}

\begin{document}
\maketitle

\begin{abstract}
In regions with heavy snowfall, the living environment is becoming a serious problem due to heavy snow accumulation. 
A road heating is an electrical device which
promotes snow melting by burying a heating cable as a thermal source underground in such regions. 
When integrating the road heating into power distribution systems, 
we need to optimize the flow of {{electric power}} by appropriately integrating distributed power sources and conventional power distribution equipment. 
In this paper, 
we introduce a battery storage to the power distribution system including road heating, 
and extend the predictive switching control of the systems due to the authors' previous study 
to the case where battery storage is installed. 
As a main result, we propose a predictive switching control that utilizes photovoltaic (PV) power generation and surplus power stored in the battery storage effectively, 
and achieves the reduction of distribution loss, attenuation of voltage fluctuation, and efficient snow melting, simultaneously. 
We verify the effectiveness of the application of battery storage through numerical simulation using actual time series data of weather conditions and active power of the PV power generation and load. 
\end{abstract}

\section{Introduction}


In heavy snowfall regions, 
there has long been a serious problem of great inconvenience and danger to people's lives, for example falling on sidewalks, heavy workload due to snow removal, traffic congestion and etc~{\cite{road}\cite{stat}}. 
A road heating is one of the electrical devices which promotes snow melting by burying a thermal source underground, and raises the temperature of the ground surface~
\cite{eu01}. 
However, the road heating consumes a large amount of electric power 
when we integrate {{it}} 
into the power distribution system~\cite{dist1}\cite{powe1}.
The flow of {{electric power}} should be optimized 
by properly integrating distributed power sources, e.g. photovoltaic power generation (PV), and electric vehicles (EVs) with conventional power distribution equipment 
for an appropriate operation of power distribution systems. 


Against the problem in the above paragraph, 
the authors of this paper~\cite{nolta} have derived a mathematical model of the power distribution system including road heating, 
and of snow melting on the ground surface via thermal diffusion in the underground based on the nonlinear ordinary differential equation (ODE) model~\cite{arxiv}\cite{haiden}. 
Based on this model, the authors proposed a switching predictive control 
of the power distribution system including road heating~\cite{sice}. 
The proposed control of \cite{sice} achieves reduction of power 
distribution losses, attenuation of voltage fluctuation, and efficient snow melting, simultaneously. 
However, the supply of {{active power}} from the PV power generation equipment is limited to the distribution system only. 
There is a potential for improvement from a view point of power flow optimization. 
In \cite{esaim}, a 2-dimensional flat ground surface is considered as the target of snow melting, and {{ForestGreen}{the}} 
melting is occurred by heat generation from thermal pipes. 
This study is different from the framework of an electrical heating cable as part of an {{electric power}} distribution system as considered in this paper.


In this paper, we extend the results of the previous study~\cite{sice} in the following two points from the observations given in the above paragraph, .
\begin{itemize}
\item
A battery storage is introduced to the considered power distribution system 
to make effective use of surplus power generated by the PV power generation in the past.
\item
We introduce a switching to the PV power generation equipment 
for a more efficient operation of the power distribution system which does not depend on transmission from the grid. 
\end{itemize}
Based on the above extension, 
we give a design of predictive switching control of the {{power}} distribution system 
including the switching of the PV power generation equipment and battery storage as a main result. 
Note that the authors of this paper have reported preliminary result in the reference~\cite{mscs} toward the objective explained at the beginning of this paragraph. 
The results of this paper provide a mathematically more accurate description of the mathematical model of the considered power distribution system than the reference~\cite{mscs}. In addition, this paper provides an additional case on the weather conditions for the simulation verification, 
and give comparisons with the result for the case without the battery storage in the authors' previous study~\cite{sice}.


The outline of this paper is described as follows. 
In Section~\ref{sec:mode}, 
we introduce a mathematical model of the considered power distribution system including 
{{the}} switching of the PV power generation equipment and battery storage. 
As a main result, we give a switching predictive control for the system based on an {{the}} evaluation {{functions}} to simultaneously achieve reduction of distribution losses, attenuation of voltage impact, and efficient snow melting in Section~\ref{sec:swit}. 
Finally, we verify an effectiveness of the proposed switching predictive control 
and the application of the battery storage 
via numerical simulation using actual 
data of weather conditions and active power of PV {{power generation}} 
and load in Section~\ref{sec:nume}. 

\section{Modeling of Power Distribution System}
\label{sec:mode}

In this section, we develop the distribution system including road heating considered in the previous studies due to the authors~\cite{nolta}\cite{sice} to a mathematical model for the case including {{the}} switching of the discharging of {{the}} PV power generation equipment and battery storage. 
In Subsection~\ref{sec:enti}, we describe the overall power distribution system considered in this paper. 
We provide the mathematical models of the voltage and temperature distribution of the underground power distribution line and heating cable that constitute the distribution system in Subsections~\ref{sec:mode_dist} and \ref{sec:mode_heat}, respectively. 
Furthermore, in Subsection~\ref{sec:mode_swit}, we introduce the mathematical models of the branch points, switches and transformers that interconnect them. Finally, in Subsection~\ref{sec:mode_ther}, we describe the mathematical models of thermal diffusion in the underground and the snow melting on the ground surface. 
The details of the mathematical model of the overall system is summarized in the authors' previous studies~\cite{nolta}\cite{sice}. 

\subsection{Overall setup of system}
\label{sec:enti}

In this subsection, we describe the overall setup of the {{considered}} power distribution system including road heating.

\begin{figure}[htb]
\centering
\includegraphics[width=0.95\columnwidth]{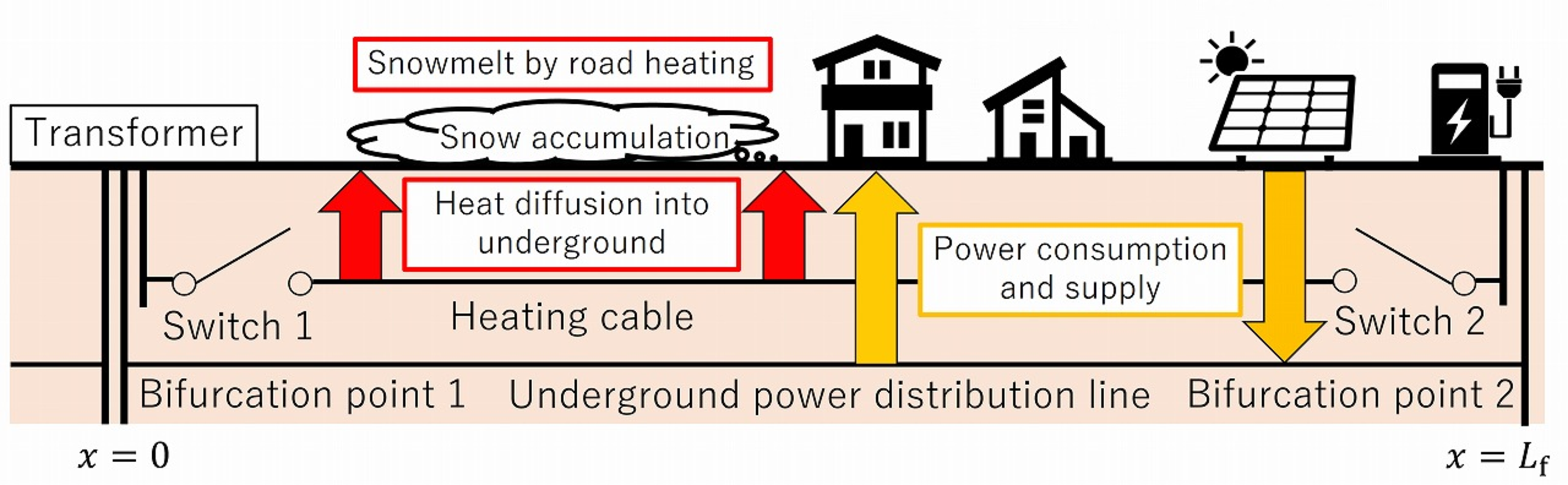}
\caption{Considered power distribution system}
\label{zentai}
\end{figure}

\begin{figure}[htb]
\centering
\includegraphics[width=0.95\columnwidth]{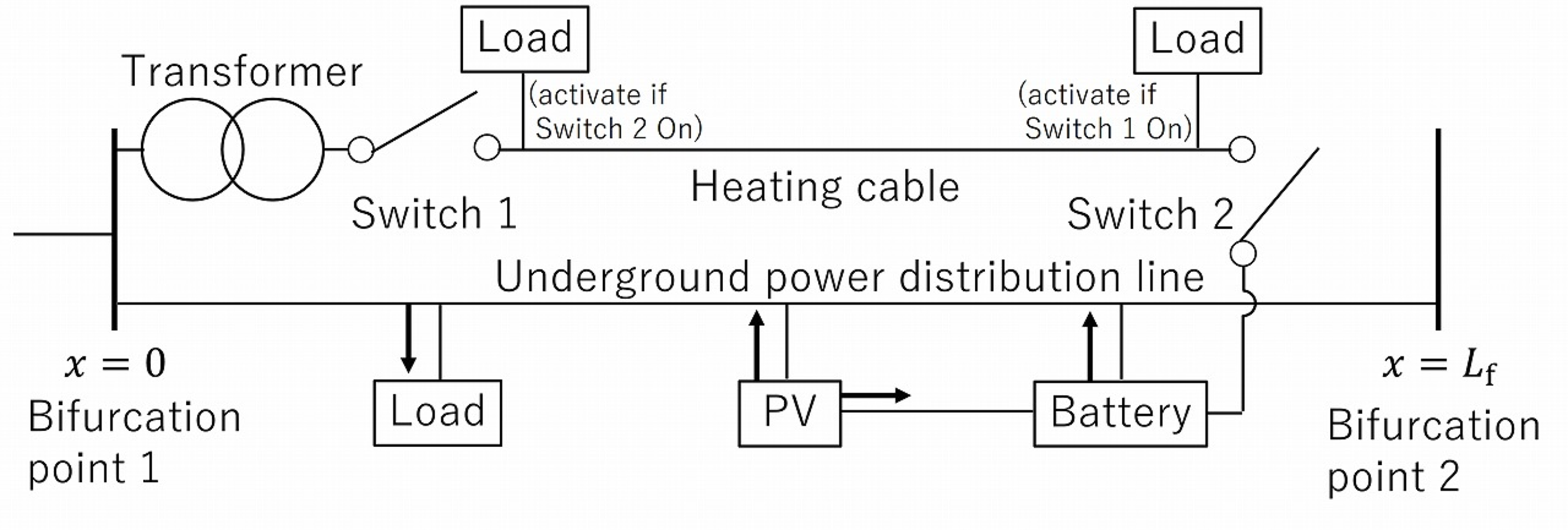}
\caption{{Single-line diagram of the considered power distribution system}}
\label{haidenzentai}
\end{figure}

{{We}} show an overview of the system in Fig.~\ref{zentai}. 
In the system, {{electric power}} is supplied from the underground power distribution line to the load electrical equipment (residences, EVs, lamps, and etc.) and from the PV power generation equipment to the underground power distribution line. 
Assuming a realistic burial location, 
a heating cable is buried in parallel with the underground power distribution line in the horizontal direction. 
{{We also show a single-line diagram of the power distribution system 
in Fig.~\ref{haidenzentai}.
The branches of the underground power distribution line and the heating cable at the starting point and terminal are denoted as Bifurcation points~1 and 2, respectively. 
Then, the electrical equipment connected to the underground power distribution line and heating cable is explained by the following~(i)-(iii). 
\begin{itemize}
\item[(i)]
Two switches are connected at both terminals to control the voltage distribution over the heating cable. 
When Switch~1 is turned \texttt{On}, the {{active power}} is transmitted from the starting end of the heating cable. On the other hand, the power is transmitted from the battery storage if Switch~2 is turned \texttt{On}. 
\item[{{(ii)}}]
{{
The terminal of the heating cable connects to either Bifurcation point~1 or 2 
in response to the switching explained above. 
As shown in Fig.~\ref{haidenzentai}, in order to allow a current to flow through the heating cable due to this switching, we suppose that a load, which is activated by the switching, is connected to the terminal that has been switched.}}
\item[(iii)]
The power distribution system is equipped with a battery storage, 
which allows surplus electric power generated by the PV power generation equipment 
to be charged to the battery storage via switching. 
In addition, the stored power can be supplied to the underground power distribution line and heating cable through switching of discharging from the battery storage.
\end{itemize}}}

{{The heat generated in the cable of the distribution line is transferred within the cable, and diffused into the underground. 
The Joule heat propagating to the ground surface is used to melt snow. 
No {{load electrical equipment (e.g. residences, EVs, lamps, and etc.)}} or PV power generation equipment is connected at the terminal of the underground power distribution line and the open ends of the heating cable.
We suppose that the Joule heat generated by the underground power distribution line is sufficiently small compared to the heat due to the heating cable.}}

\subsection{Voltage distribution on underground power distribution line~\cite{arxiv}\cite{haiden}}
\label{sec:mode_dist}

We consider the three-phase underground power distribution line where the line extends in a straight line 
from the transformer to the load electrical equipment and PV {{power}} generation equipment 
between both bifurcation points 
as shown in Fig.~\ref{haidenzentai}. 
We also suppose that there are no voltage regulation devices such as transformers 
on the intermediate positions over the lines. 
The horizontal spatial variable on the underground power distribution line is defined as $x$~[m]. 
The starting point and terminal of the line correspond to $x=0$~[m] and $x=L_{\rm f}$~[m], $L_{\rm f}>0$, respectively. 

\subsubsection{Voltage distribution}

We describe the voltage distribution of the underground power distribution line, which is assumed to be in three-phase equilibrium, by the ODEs 
with position~$x$~[m] as an independent variable, for an arbitrarily fixed time~$t$~[min]. 
We denote the voltage phasor for single-phase at position~$x$ and time~$t$ by $v_{\rm e}(x;t)e^{j\theta_{\rm e}(x;t)}$, 
where $j$ is the imaginary unit. 
Then, $\theta_{\rm e}(x;t)$~[rad] and $v_{\rm e}(x;t)$~[V] correspond to the voltage phase and amplitude, respectively. 
The spatial variation on the distribution voltage profile 
is described 
by the nonlinear ODE model~\cite{arxiv}\cite{haiden}: 
\begin{align}
\nonumber
\frac{\mathrm{d}\theta_{\rm e}(x;t)}{\mathrm{d}x}&=-\frac{s_{\rm e}(x;t)}{v_{\rm e}(x;t)^2},\\
\nonumber
\frac{\mathrm{d}v_{\rm e}(x;t)}{\mathrm{d}x}&=w_{\rm e}(x;t),\\
\label{ode s}
\frac{\mathrm{d}s_{\rm e}(x;t)}{\mathrm{d}x}&=\frac{b_{\rm e}p_{\rm e}(x,t)-g_{\rm e}q_{\rm e}(x,t)}{g_{\rm e}^2+b_{\rm e}^2},\\
\label{ode w}
\frac{\mathrm{d}w_{\rm e}(x;t)}{\mathrm{d}x}&=\frac{s_{\rm e}(x;t)^2}{v_{\rm e}(x;t)^3}-\frac{g_{\rm e}p_{\rm e}(x,t)+b_{\rm e}q_{\rm e}(x,t)}{(g_{\rm e}^2+b_{\rm e}^2)v_{\rm e}(x;t)},
\end{align}
where 
$s_{\rm e}(x,t)$~[V${}^2$/m] 
and $w_{\rm e}(x,t)$~[V/m] represent 
the supplemental variable and voltage gradient at position~$x$ and time~$t$, respectively. 
In (\ref{ode s}) and (\ref{ode w}), $p_{\rm e}(x,t)$~[W/m] and $q_{\rm e}(x,t)$~[Var/m] are active and reactive powers at position~$x$ and time~$t$ of the underground power distribution line, respectively{{, which are transmitted from or to the PV power generation equipment, load electrical equipment, and battery storage.}} 
If $p_{\rm e}(x,t)>0$ holds, $p_{\rm e}(x,t)$ denotes the active power, i.e. the supply, flowing to the distribution line at $x$. 
On the other hand, if $p_{\rm e}(x,t)<0$ holds, $p_{\rm e}(x,t)$ corresponds to the active power consumption flowing from the line at $x$. 
The same inequality holds for the reactive power $q_{\rm e}(x,t)$. 
Due to the assumptions made at the terminal of the underground power distribution line,  
$p_{\rm e}(L_{\rm f}, t)=0$ and $q_{\rm e}(L_{\rm f},t)=0$ hold.
The constants~$g_{\rm e}$~{{[m/$\Omega$]}} and $b_{\rm e}$~{{[m/$\Omega$]}} are the conductance and susceptance per unit length at position~$x$ of the line. 

\subsubsection{Boundary condition}
\label{tityuhaidensennokyokaijyoken}

The voltage phase and amplitude of the underground power distribution line satisfy the following boundary conditions at each point:
\begin{align}
\mbox{\underbar{Bifurcation point~1}: }&\theta_{\rm e}(0,t)=\theta_1, \ 
v_{\rm e}(0,t)={v_1}, 
\label{bifurcation1}\\
\mbox{\underbar{Bifurcation point~2}: }&\theta_{\rm e}(L_{\rm f},t)=\theta_2(t), \ 
v_{\rm e}(L_{\rm f},t)=v_2(t),
\label{bifurcation2}
\end{align}
In (\ref{bifurcation1}), $\theta_1$ and $v_1$ denote the constant voltage phase and amplitude in Bifurcation point~1, respectively, which correspond to the reference voltage phase and amplitude for the line at $x=0$. 
Moreover, $\theta_2(t)$ and $v_2(t)$ are the voltage phase and amplitude in Bifurcation point~2 at time~$t$, respectively, in (\ref{bifurcation2}).

\subsection{Voltage and temperature distribution of heating cable~\cite{nolta}}
\label{sec:mode_heat}

We assume that the heating cable is single-phase two-wire system. 
The heating cable extends in a straight line from the transformer at the starting point, 
to the terminal, 
with the switches connected at both ends. 
The horizontal spatial variable of the heating cable is $x$~[m] in common with underground power distribution line. 
The vertical spatial variable~$y$~[cm] is the depth 
from the ground surface to the heating cable. 
This implies that $y=0$~[cm] and $y=D_{\rm f}$~[cm] correspond to the ground and the surface, respectively. 
The length of the switches is negligible because it is supposed sufficiently short compared to the underground power distribution line and heating cable. 
No voltage regulation devices 
are installed on the intermediate positions over the cable 
 similarly to the underground power distribution line. 

\subsubsection{Voltage distribution}
\label{sec:volh}

We denote the voltage phasor for single-phase at position~$x$ and time~$t$ by $v_{\rm h}(x;t)e^{j\theta_{\rm h}(x;t)}$, 
where $\theta_{\rm h}(x;t)$~[rad] and $v_{\rm h}(x;t)$~[V] are the voltage phase and amplitude, respectively. 
Similarly to the underground power distribution line, 
%
the spatial variation of the voltage distribution over the heating cable, which is assumed to be single-phase two-wire system,  is described by the nonlinear ODE model~\cite{nolta}:
\begin{align}
\frac{\mathrm{d}\theta_{\rm h}(x,t)}{\mathrm{d}x}&=-\frac{s_{\rm h}(x,t)}{v_{\rm h}(x,t)^2},
\nonumber \\
\frac{\mathrm{d}v_{\rm h}(x,t)}{\mathrm{d}x}&=w_{\rm h}(x,t),
\nonumber \\
\label{ode sh}
\frac{\mathrm{d}s_{\rm h}(x,t)}{\mathrm{d}x}&=\frac{b_{\rm h}p_{\rm h}(x,t)-g_{\rm h}q_{\rm h}(x,t)}{g_{\rm h}^2+b_{\rm h}^2},\\
\label{ode wh}
\frac{\mathrm{d}w_{\rm h}(x,t)}{\mathrm{d}x}&=\frac{s_{\rm h}(x,t)^2}{v_{\rm h}(x,t)^3}-\frac{g_{\rm h}p_{\rm h}(x,t)+b_{\rm h}q_{\rm h}(x,t)}{(g_{\rm h}^2+b_{\rm h}^2)v_{\rm h}(x,t)},
\end{align}
where 
$s_{\rm h}(x,t)$~[V${}^{2}$/m] 
and $w_{\rm h}(x,t)$~[V/m] represent 
the supplemental variable and voltage gradient at position~$x$ and time~$t$, respectively. 
In (\ref{ode sh}) and (\ref{ode wh}), $p_{\rm h}(x,t)$~[W/m] and $q_{\rm h}(x,t)$~[Var/m] are active and reactive powers at position~$x$ and time~$t$ of the heating cable, respectively. 
These powers satisfy the same properties that described for the underground power distribution line in the previous section. 
Moreover, $g_{\rm h}$~{{[m/$\Omega$]}} and $b_{\rm h}$~{{[m/$\Omega$]}} are the conductance and susceptance per unit length at position~$x$ of the cable, respectively. 

\subsubsection{Boundary condition}
\label{sec:boun}

We describe the boundary conditions of the voltage phase and amplitude, supplemental variable, and voltage gradient at positions~$x=0$ and $x=L_{\rm f}$ over the heating cable. 
The boundary conditions 
are described as follows corresponding to the switching patterns: 
\begin{align*}
& \mbox{\underline{Switches~1 and 2: {\tt Off}}: } \\
& 
\theta_{\rm h}(0;t)=0, \ v_{\rm h}(0;t)=0, \\ 
& s_{\rm h}(0;t)=0, \ w_{\rm h}(0;t)={0},\ 
\theta_{\rm h}(L_{\rm f};t)=0, \ v_{\rm h}(L_{\rm f};t)=0, \\
& s_{\rm h}(L_{\rm f};t)=0, \ w_{\rm h}(L_{\rm f};t)=0,\\
& \mbox{\underline{Switch~1: {\tt On}}: } \\
& 
\theta_{\rm h}(0;t)=\theta_1, \ 
v_{\rm h}(0;t)=\frac{1}{{a_{\rm 1}}}v_1, \\
& 
s_{\rm h}(L_{\rm f};t)=0, \ 
w_{\rm h}(L_{\rm f};t)=0,\\
& \mbox{\underline{Switch~2: {\tt On}}: } \\
& 
s_{\rm h}(0;t)=0, \ 
w_{\rm h}(0;t)=0,\\
& 
\theta_{\rm h}(L_{\rm f};t)=\theta_1, \ 
v_{\rm h}(L_{\rm f};t)=\frac{1}{{a_{\rm 1}}}v_1.
\end{align*}
In the above conditions, 
$a_{\rm 1}$ is the transformer ratio of the single phase transformer. 
{{Moreover, the active power~$p_{\rm h}(x_{\rm load},t)$ is transmitted to the heating cable 
due to the load connected to the terminal of the cable corresponding to the switching as described in Section~\ref{sec:enti}. 
We define the connection position of the load which corresponds to the terminal as follows: 
\begin{align}
\begin{cases}
\text{No load connected} & \text{(Switches~1 and 2: {\tt Off})}\\
x_{\rm load}\fallingdotseq L_{\rm f} & \text{(Switch~1: {\tt On})}\\
x_{\rm load}\fallingdotseq 0 & \text{(Switch~2: {\tt On}). }
\end{cases}
\label{hload}
\end{align}}}

\subsubsection{Thermal distribution of heating cable surface}
\label{sec:ode_surf}

At the surface of the heating cable, the transient spatial and temporal variation of the temperature satisfies the following ODE: 
\begin{align*}
& \frac{\mathrm{d}\delta_{\rm surf,diff}(t;x)}{\mathrm{d}t}\\
& =\frac{\Gamma_{\rm h}(x,t)\times10^{-2}-q_{\rm r}-\gamma_{\rm cable}\delta_{\rm surf,diff}(t;x)}{C_{\rm cable}}
\end{align*}
from \cite{sug:eval1}\cite{ino:effe1}, 
where $\delta_{\rm surf,diff}(t;x)$ [${}^{\circ}{\rm C}$] is the difference of the temperature between the surface 
and the ambient surroundings.
In addition, 
$C_{\rm cable}$ [J/cm$\cdot {}^{\circ}{\rm C}$] is the heat capacity of {the cable}, 
$q_{\rm r}$ [W/cm] is the radiative cooling 
supposed to be constant, 
and $\gamma_{\rm cable}$~[W/m$\cdot$K] is the contact heat transfer coefficient per unit length between the cable surface and soil. 
From \cite{soudenrosu}, $\Gamma_{\rm h}(t;x)$~[W/m] is the power distribution loss (Joule heat) of the heating cable at position~$x$ and time~$t$:
\begin{align}
\label{gammah}
\Gamma_{\rm h}(t;x)&=g_{\rm h}\left(w_{\rm h}(t;x)^2+\frac{s_{\rm h}(t;x)^{2}}{v_{\rm h}(t;x)^{2}}\right).
\end{align}
The cable surface temperature $\delta_{\rm 
surf}(t;x)$~[${}^\circ$C] at position~$x$ and time 
$t$ is given by $\delta_{\rm surf}(t;x) = \delta_{\rm surf,diff}(t;x) + \delta_{\rm soil}(D_{\rm f},t;x)$, 
where $\delta_{\rm soil}(D_{\rm f},t;x)$~[${}^\circ$C] is the temperature of the soil at position~$x$ and depth~$y=D_{\rm f}$. 
Based on the surface temperature and contact heat transfer~\cite{itijigenkyokai}, 
the boundary condition at the surface 
is given by 
\begin{align*}
-\lambda_{\rm tran}\frac{\mathrm{d} \delta_{\rm soil}(D_{\rm f},t;x)}{\mathrm{d} y}=\beta_{\rm cable}(\delta_{\rm surf}(x,t)-\delta_{\rm soil}(D_{\rm f},t;x)),
\end{align*}
where $\lambda_{\rm tran}$~[W/m$\cdot$K] and $\beta_{\rm cable}$~[W/m${}^{2}\cdot$K] are the thermal conductivity of the soil and the heat transfer coefficient
between heating cable and soil, respectively. 

\subsection{Switches}
\label{sec:mode_swit}

Since the voltages at Transformer~B and the terminal $(x=L_{\rm f})$ of the underground power distribution line are required to be equal, if one switch is {\tt On}, the other switch is {\tt Off}. 
Moreover, if the heating cable is not used, both switches are {\tt Off}.
We define binary variable~$\sigma_{i}(t) \in \{0,1\}$ $(i=0,1,2)$ by
\begin{align*}
\sigma_{0}(t)&:=
\begin{cases}
1 & \mbox{(Switches~1 and 2: {\tt Off})} \\ 
0 & \mbox{(Otherwise), } \\ 
\end{cases}\\ 
\sigma_{1}(t)&:=
\begin{cases}
1 & \mbox{(Switch~1: {\tt On}, Switch~2: {\tt Off})} \\ 
0 & \mbox{(Otherwise), } \\ 
\end{cases} \\ 
\sigma_{2}(t)&:=
\begin{cases}
1 & \mbox{(Switch~1: {\tt Off}, Switch~2: {\tt On})} \\ 
0 & \mbox{(Otherwise)}. \\ 
\end{cases}
\end{align*}
These variables can be regarded as the control input to the power distribution system. 
We refer to the integer~$i$ as the index of the control input~$\sigma_{i}(t)$. 
{{One of the these variables must have the value~$1$ from the definitions of the variables. This implies that the variables satisfy the constraint~$\sigma_{0}(t)+\sigma_{1}(t)+\sigma_{2}(t)=1$.}}


We assume that the PV power generation equipment can transmit the generated {{active power}} to either underground power distribution line or battery storage. 
{{For simplicity, we do not take into account the power generation efficiency and conversion efficiency in relation to solar radiation.}} 
We define the binary variable~$\sigma_{{\rm pv}}(t) \in \{0,1\}$ related to the power transmission for the equipment 
by corresponding to the direction of the power transmission as follows: 
\begin{align*}
\sigma_{{\rm pv}}(t)&:=
\begin{cases}
1 & \mbox{(Underground power distribution line)} \\ 
0 & \mbox{(Battery storage)}. \\ 
\end{cases} 
\end{align*}


In the battery storage, it is assumed that the active power is supplied only from the PV power generation equipment. 
We assume that a certain amount of power can be transmitted to either the heating cable or the underground power distribution line, respectively. 
We also allow for the possibility of not transmitting power to either of them. 
{{Furthermore, the charging and discharging efficiency of the battery storage can be ignored for the same reason as the PV power generation.}}
The binary variable~$\sigma_{{\rm battery},i}(t) \in \{0,1\}$ $(i=0,1,2)$ for the transmission of power from the battery storage is defined by the following equality corresponding to the direction of the power transmissions 
as the control input to the power distribution system:
\begin{align*}
\sigma_{{\rm battery},0}(t)&:=
\begin{cases}
1 & \mbox{(No power transmission)} \\ 
0 & \mbox{\begin{tabular}{@{\,}l}(Underground power distribution line\\[-1mm]
 or heating cable),\\
 \end{tabular}} \\ 
\end{cases}\\ 
\sigma_{{\rm battery},1}(t)&:=
\begin{cases}
1 & \mbox{(Underground power distribution line)} \\
0 & \mbox{\begin{tabular}{@{\,}l}
(Heating cable\\[-1mm]
or no power transmission),\\ 
\end{tabular}} \\ 
\end{cases} \\ 
\sigma_{{\rm battery},2}(t)&:=
\begin{cases}
1 & \mbox{(Heating cable)} \\ 
0 & \mbox{\begin{tabular}{@{\,}l}(Underground power distribution line\\[-1mm]
or no power transmission),\\
\end{tabular}} \\ 
\end{cases}
\end{align*}
{{Since one of these variables is always equal to $1$ and the other $0$ similarly to the binary variables of the switches of the heating cable, 
$\sigma_{{\rm battery},i}(t)$ $(i=0,1,2)$ satisfy the constraint~$\sigma_{{\rm battery},0}(t)+\sigma_{{\rm battery},1}(t)+\sigma_{{\rm battery},2}(t)=1$.}}

\subsection{Thermal Diffusion in Underground and Snow Volume~\cite{nolta}}
\label{sec:mode_ther}

\subsubsection{Thermal diffusion}
\label{sec:pde_soil}

Throughout this paper, we assume that there is no horizontal thermal diffusion, 
and that the diffusion occurs only in the vertical direction in the underground. 
We define the underground temperature of the soil at position~$x$,~depth $y$ and time~$t$ by $\delta_{\rm soil}(y,t;x)$~[${}^\circ$C]. 
The spatial and temporal variation of the temperature is described by the following standard thermal conduction equation: 
\begin{align*}
\frac{\partial \delta_{\rm soil}(y,t;x)}{\partial t}&=\alpha_{\rm soil}\frac{\partial^2 \delta_{\rm soil}(y,t;x)}{\partial y^2},
\end{align*}
where $\alpha_{\rm soil}$~[m${}^2$/s] is thermal diffusivity of soil, respectively.
The boundary condition at the ground surface 
is given by the thermal conduction equation 
with the thermal exchange from the snow accumulation~\cite{itijigenkyokai}: 
\begin{align*}
-\lambda_{\rm tran}\frac{\mathrm{d} \delta_{\rm soil}(0,t;x)}{\mathrm{d} y}=\beta_{\rm ground}(\delta_{\rm soil}(0,t;x)-\delta_{\rm snow}),
\end{align*}
where 
$\beta_{\rm ground}~\mathrm{[W/m\cdot K]}$ 
is the contact heat transfer coefficient per unit length between the ground surface and accumulation of snow.
Moreover, {$\delta_{\rm snow}$~[${}^\circ$C]} is the constant snow temperature on the ground surface. 

\subsubsection{Snow melting}
\label{sec:ode_snow}

We suppose that the snow melting occurs due to the radiant heat from the sunlight 
and the Joule heat of the heating cable for simplicity. 
The spatial and temporal variation of the snow volume~$h_{\rm snow}(x,t)$~[mm] is described by 
by the following equation~\cite{netusyusi}: 
\begin{align}
\label{suitoryo}
\frac{\mathrm{d}h_{\rm snow}(t;x)}{\mathrm{d}t}&=-\frac{a_{\rm snow}}{d_{\rm snow}}(\mu_1(x,t)+\mu_2(x,t))+f_{\rm snow}(t).
\end{align}
where $h_{\rm snow}(x,t)$~[mm] is the snow volume and $f_{\rm snow}(t)$~[mm/min] is the snowfall per 1~minute. 
Moreover, 
$d_{\rm snow}$~[g/cm${}^{3}$] is the density of snow volume, 
and $a_{\rm snow}$ is  the unit conversion factor from the unit~W/m${}^2$ to the unit~mm/min. 
Finally, $\mu_1(x,t) $~[W/m${}^2$] represents the snow melting due to the sunlight at position~$x$ and time~$t$. 
The snow melting due to the influence of the sunlight is given by $\mu_1(x,t)=\phi_{\rm r}(t)+\phi_{\rm s}(x,t)+\phi_{\rm l}(x,t)$, 
where $\phi_{\rm r}(t) $~[W/m${}^2$] is the net radiation flux at time~$t$, which is assumed to be independent on position~$x$. 
{Moreover, $\phi_{\rm s}(x,t) $~[W/m${}^2$] and $\phi_{\rm l}(x,t) $~[W/m${}^2$] denote the sensible and latent heat fluxes at position~$x$ and time~$t$, respectively.} 
In addition, $\mu_2(x,t) $~[W/m${}^2$] represents the snow melting due to the Joule heat of the cable at position~$x$ and time~$t$. 
Finally, in (\ref{suitoryo}), the snow melting due to the Joule heat of the heating cable is expressed as 
$\mu_2(x,t)=\beta_{\rm ground}(\delta_{\rm soil}(0,t;x)-\delta_{\rm snow})$. 

\section{Switching Predictive Control} 
\label{sec:swit}

In this section, we explain the predictive control by switches connected to the heating cables and the predictive control of the PV power generation equipment and the battery storage as the main result of this paper. 
At first, in Subsection~\ref{switchnosettei}, we describe the overall design of the proposed predictive switching control. Next, in Subsection~\ref{sec:heat}, the evaluation function and optimal switching patterns are introduced for the switching of the heating cable. Moreover, a similar formulation is given for the PV power generation equipment in Subsection~\ref{sec:pv}. Finally, in Subsection~\ref{sec:battery}, a mathematical description is provided with respect to the optimal discharging of the battery storage.

\subsection{Design of Switching Predictive Control}
\label{switchnosettei}

\begin{figure}[htb]
\centering
\includegraphics[width=0.92\columnwidth]{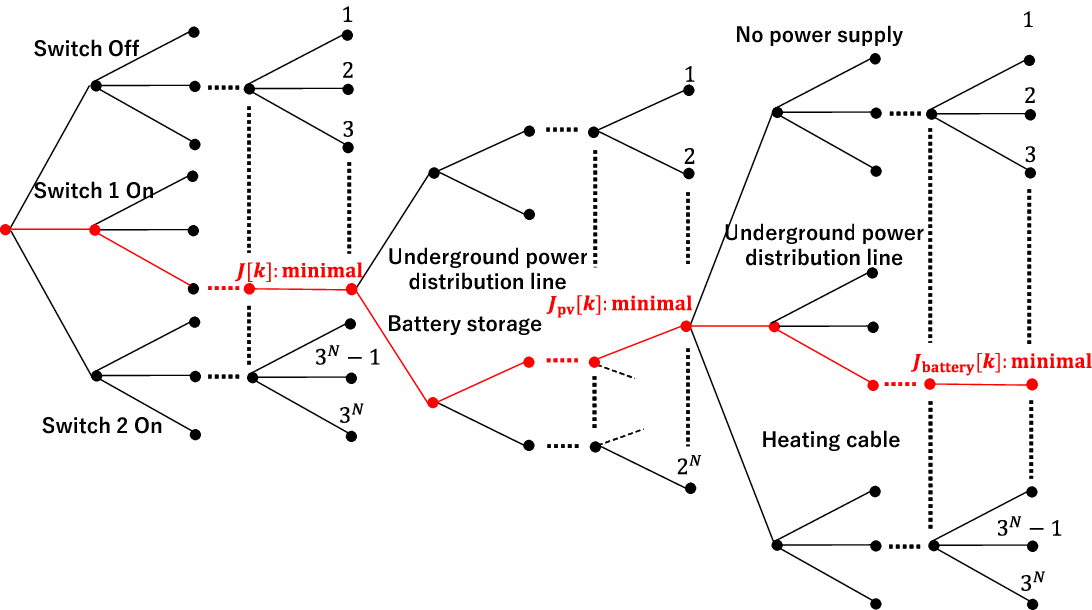}
\caption{Selection tree of switches}
\label{sentakubunki}
\end{figure}

In this subsection, we explain the overall process for the predictive switching control of the power distribution system, the PV power generation equipment, and predictive control of battery storage. 
The optimal control inputs are computed according to Steps~0-5 below. 
Note that Step~0 and Steps~1-4 correspond to the preliminary and during operation of the predictive control, respectively. 
We also add Steps~3 and 4 to the predictive switching control of the previous study~\cite{sice} due to the authors. 
\begin{itemize}
\item
\underline{Step~0}: 
We define the minimum switching time between the switching and the power transmission from the battery storage as $T_{\rm mini}$~[min]. 
Moreover, we define the predictive horizon as $T_{\rm pred}$~[min]. 
We prepare the time series data at each time ~$t=kT_{\rm mini}$ $(k=0,1,2,\cdots)$ for the residential load, PV power generation, solar radiation, snowfall, temperature, wind speed, and snow accumulation 
as inputs for the predictive simulation.
\item
\underline{Step~1}: 
We define $N:=\frac{T_{\rm pred}}{T_{\rm mini}}$ as the total number of times at which switching can occur. 
This number corresponds to the sum of all $3^{N}$~switching patterns, $2^{N}$~PV transmission patterns, and $3^{N}$~discharging patterns of the battery storage.  
For all possible 
{{$3^N \times 2^N \times 3^N = 18^{N}$}}~patterns, we perform the simulations from $t=kT_{\rm mini}$ to $t+T_{\rm pred}=(k+N)T_{\rm mini}$ based on the mathematical model of the distribution system given in Section~\ref{sec:mode}, 
\item
\underline{Step~2}: 
Based on the simulation in Step~1, 
we compute the value of the evaluation function~$J[k]$ for the possible $3^{N}$~switching patterns~$\vb*{\mathbf{\sigma}}_{\rm d}[k]$ at $t=kT_{\rm mini}$. 
We then compare the $3^{N}$~values of the functions. 
We choose the optimal switching pattern~$\vb*{\mathbf{\sigma}}_{\rm d}[k]$ 
that has the smallest value at $t=kT_{\rm mini}$. 
We explain the detailed procedure of the design of the optimal switching pattern in Subsection~\ref{sec:heat}. 
\item
\underline{Step~3}: 
Based on the simulation and the optimal switching pattern~$\vb*{\mathbf{\sigma}}_{\rm d}^{*}[k]$ computed in Steps~1 and 2, respectively, 
we compute the evaluation function for the $2^{N}$~possible transmission patterns of the PV power generation equipment at $t=kT_{\rm mini}$. 
We compute the value of the evaluation function~$J_{\rm pv}[k]$ for $\vb*{\mathbf{\sigma}}_{\rm d,pv}[k]$ of the transmission patterns of the PV power generation equipment at $t=kT_{\rm mini}$. 
Then, by comparing the $2^{N}$ values of the evaluation functions, 
we find the optimal transmission pattern~$\vb*{\mathbf{\sigma}}_{\rm d,pv}[k]$ of the PV power generation equipment whose value is the minimum at $t=kT_{\rm mini}$. 
The detailed procedure of this step is explained in Subsection~\ref{sec:pv}. 
\item
\underline{Step~4}: 
Based on the predictive simulation in Step~1, the optimal switching pattern~$\vb*{\mathbf{\sigma}}_{\rm d}^{*}[k]$ in Step~2, and the optimal transmission pattern~$\vb*{\mathbf{\sigma}}_{\rm d, pv}[k]$ in Step~3, 
we compute the values of the evaluation function~$J_{\rm battery}[k]$ for the $3^{N}$~battery storage transmission pattern~$\vb*{\mathbf{\sigma}}_{\rm d,battery}[k]$ at $t=kT_{\rm mini}$ for the evaluation function~$J_{\rm battery}[k]$. 
In this case, the $3^{N}$~values of the evaluation function are compared, 
and the battery transfer pattern~$\vb*{\mathbf{\sigma}}_{\rm d,battery}[k]$ with the minimum value at $t=kT_{\rm mini}$ is selected as the optimal power transfer pattern~$\vb*{\mathbf{\sigma}}_{\rm d,battery}^{*}[k]$. 
We explain the detailed procedure of the design of the optimal switching pattern in Subsection~\ref{sec:battery}. 
\item
\underline{Step~5}: 
For the optimal switching pattern and the optimal transmission pattern computed in the previous steps, 
the control inputs are continuously applied from $t=kT_{\rm mini}$ to $t+T_{\rm mini}=(k+1)T_{\rm mini}$. 
Then, we choose the switching pattern which has the smallest value of the 
evaluation function as the optimal switching pattern.
Fig.~\ref{sentakubunki} illustrates the overall flow of the switching and power transmission selection from Step~2 to Step~4. 
\end{itemize}

\subsection{Optimal Switching of Heating Cable}
\label{sec:heat}

In this subsection, we give the evaluation function that determines the switching 
on the heating cable. 
We define the evaluation function~$J[k]$ at $t=kT_{\rm mini}$ by
\begin{align*}
J[k]:=  & w_{\rm loss}P_{\rm loss}[k]+w_{\rm fluc}V_{\rm fluc}[k]
+w_{\rm snow}S_{\rm snow}[k] 
\nonumber \\
& +w_{\rm cost}M_{\rm cost}[k].
\end{align*}
The exact definitions and physical meaning of each term 
are explained in the remainder of this section.
\begin{itemize}
\item
\underline{The first term}: 
The function~$P_{\rm loss}{[k]}$~[W$\cdot$min] corresponds to the distribution loss of overall system in the time interval~$[t, t+T_{\rm pred})$, and $w_{\rm loss}$ is the weighting coefficient corresponding the loss.
This function is defined 
by the distribution loss of the underground power distribution line and heating cable:
\begin{align}
P_{\rm loss}{[k]}&:=  {\sum_{l=1}^{N}\int_{(k+l-1)T_{\rm mini}}^{(k+l)T_{\rm mini}}}\int_{0}^{L_{\rm f}}\Gamma(x;t) \mathrm{d}x\mathrm{d}t, 
\nonumber \\
\Gamma(x;t)&:=\Gamma_{\rm e}(x;t)+\Gamma_{\rm h}(x;t).
\label{denryokusonsitu}
\end{align}
In (\ref{denryokusonsitu}), $\Gamma_{\rm e}(x;t)$~[W/m] and $\Gamma_{\rm h}(x;t)$~[W/m] represent the distribution loss of the underground power distribution line and heating cable, respectively. 
Since the distribution loss of underground power distribution line is considered for three phases, 
the distribution loss for a single-phase is three times larger and is given by 
\begin{align*}
\Gamma_{\rm e}(x;t)=3g_{\rm e}\left(w_{\rm e}(x;t)^2+\frac{s_{\rm e}(x;t)^2}{v_{\rm e}(x;t)^2}\right)
\end{align*} 
from \cite{soudenrosu}, and $\Gamma_{\rm h}(x,t)$~[W/m] is given by (\ref{gammah}). 
\item
\underline{The second term}: The function~$V_{\rm fluc}{[k]}$~[V] corresponds to the evaluation of the voltage {fluctuation} of the underground power distribution line in the time interval~$[t, t+T_{\rm pred})$, and $w_{\rm fluc}$ is the weighting coefficient  in the evaluation of {the  fluctuation}.
The function~$V_{\rm fluc}[k]$ is defined by the maximum absolute values of the difference between the voltage amplitude and the reference value {$v_{1}$~[V]} for $N$~time 
intervals:
\begin{align}
\label{vhendou}
V_{\rm fluc}{[k]}&:=  \sum_{{l}=1}^{N}\max_{
{\scriptsize\begin{array}{@{\,}c@{\,}}
x\in[0,L_{\rm f}], \\
\tau \in[{(k+l-1)T_{\rm mini},(k+l)T_{\rm mini}})\\
\end{array}}} \hspace*{-8mm}
\left|v_{\rm e}(x;\tau)-v_{1}\right|.
\end{align}
\item
\underline{The third term}: The function~$S_{\rm snow}{[k]}$~[m$^2\cdot$min] corresponds to the snow volume at all points in the time interval~$[t, t+T_{\rm pred})$, and is defined by
\begin{align*}
S_{\rm snow}{[k]}:=  {\sum_{{l}=1}^{N}\int_{(k+l-1)T_{\rm mini}}^{(k+l)T_{\rm mini}}}\int_{0}^{L_{\rm f}}h_{\rm snow}(x,t) \mathrm{d}x\mathrm{d}t,
\end{align*}
where $w_{\rm snow}$ is the weighting coefficient  in the evaluation of the volume.
\item
\underline{The fourth term}: The function~$M_{\rm cost}[k]$ corresponds to the cost of operating the heating cable during the time interval~$[t, t+T_{\rm pred})$, 
and $w_{\rm cost}$ represents the weighting coefficient of its evaluation in the evaluation function. 
As we suppose that 
the owner of the heating cable, the PV power generation equipment, and the battery storage is supposed to be the same. 
the cost occurs only when Switch~1 is set to {\tt On}. 
Thus, the cost~$M_{\rm cost}[k]$ is defined as 
\begin{align*}
M_{\rm cost}{[k]}:=  {\sum_{{l}=1}^{N}\int_{(k+l-1)T_{\rm mini}}^{(k+l)T_{\rm mini}}}m_{\rm cost}(t) \mathrm{d}t.
\end{align*}
In the above equality, $m_{\rm cost}(t)$ is defined by 
\begin{align*}
m_{\rm cost}(t):= 
\begin{cases}
-p_{\rm h}(x_{\rm load},t) & \mbox{(Switch~1: {\tt On})}\\
0 & \mbox{(otherwise)},\\
\end{cases}
\end{align*}
{{where $p_{\rm h}(x_{\rm load},t) \leq 0$ is the active power of the loads of the heating cable.}} 
\end{itemize}

We define the  discrete-time vector~${\bm \sigma}_{\rm d}[k]$, 
which consists of switching patterns from $t=kT_{\rm mini}$ to $t+T_{\rm pred}=(k+N)T_{\rm mini}$, by 
\begin{align*}
{\bm \sigma}_{\rm d}[k]&:= 
\begin{bmatrix}
{\bm \sigma}_{{\rm d},0}[k]\\
{\bm \sigma}_{{\rm d},1}[k]\\
{\bm \sigma}_{{\rm d},2}[k]\\
\end{bmatrix}, \\
{\bm \sigma}_{{\rm d},i}[k]&:= 
\begin{bmatrix}
\sigma_{{\rm d},i}[k]\\
\sigma_{{\rm d},i}[k+1]\\
\vdots \\
\sigma_{{\rm d},i}[k+N]\\
\end{bmatrix}
 \ (i=0, 1, 2).
\end{align*}
We enumerate all $3^{N}$ patterns~$\vb*{\mathbf{\sigma}}_{\rm d}[k]$, and perform a numerical simulation for each of them. 
We obtain the optimal switching pattern {${\bm \sigma}_{\rm d}^{*}[k]\in \left\{0,1\right\}^{N}$} at $t=kT_{\rm mini}$ 
by solving the binary minimization problem
\begin{align*}
{\bm \sigma}{^{*}[k]}
=\underset{{\bm \sigma}[k]\in \left\{0,1\right\}^{N}} {\operatorname{argmin}} \ J[k].
\end{align*}
From the current time~{$t=kT_{\rm mini}$} to the next switching time~{$t+T_{\rm mini}=(k+1)T_{\rm mini}$},
we continue to apply the following optimal control input {$\sigma_{{\rm opt},i}(t)$ $(i=0,1,2)$} by 
\begin{align*}
\sigma_{{\rm opt},i}(t)&=\sigma_{{\rm d},i}^{*}(kT_{\rm mini}) 
, \ t \in [kT_{\rm mini},(k+1)T_{\rm mini}).
\end{align*}

\subsection{Optimal Switching of PV Power Generation Equipment}
\label{sec:pv}

In this subsection, we formulate the evaluation function that determines the destination of {{active power}} from the PV power generation equipment. 
This function is mathematically defined by 
\begin{align*}
J_{\rm pv}[k]:=  w_{\rm pvfluc}V_{\rm pvfluc}[k]
-w_{\rm stor,1}B_{\rm stor,1}[k].
\end{align*}
\begin{itemize}
\item
\underline{The first term}: 
The function~$V_{\rm pvfluc}[k]$~[V] corresponds to the evaluation of the voltage fluctuation of the underground power distribution line in the time interval~$[t, t+T_{\rm pred})$, and $w_{\rm pvfluc}$ represents the weighting coefficient in the evaluation function. 
The function~$V_{\rm pvfluc}[k]$ is defined in the same way as in 
(\ref{vhendou}). 
\item
\underline{The second term}: 
The function~$B_{\rm stor,1}[k]$ corresponds to the evaluation of the battery power in the time interval~$[t, t+T_{\rm pred})$, and $w_{\rm stor,1}$ is the weighting coefficient in the evaluation function. 
Since the owner of the PV power generation equipment and the battery storage are assumed to be the same, 
this function corresponds to smoothing for the amount of {{the active power}}~$P_{\rm battery}(t)$~[kWh] of the battery storage. 
Thus, we define $B_{\rm stor,1}[k]$ to be large when the amount of {{the active power}} in the battery storage is small.
On the other hand, we set $B_{\rm stor,1}[k]$ to be small when the amount of {{the active power}} in the battery storage is large. 
In this case, $B_{\rm stor,1}{[k]}$ can be defined by the following equality: 
\begin{align*}
B_{\rm stor,1}{[k]}:=  {\sum_{{l}=1}^{N}\int_{(k+l-1)T_{\rm mini}}^{(k+l)T_{\rm mini}}}\zeta_{\rm stor,1}(t) \mathrm{d}t, 
\end{align*}
where $\zeta_{\rm stor,1}(t)$ is defined as follows corresponding to the power transmissions: 
\begin{align*}
\zeta_{\rm stor,1}(t):= 
\begin{cases}
0 & \mbox{\begin{tabular}{@{\,}l}
(Underground power\\[-1mm] 
distribution line)\\ 
\end{tabular}}\\
\frac{1}{P_{\rm battery}(t)} & \text{(Battery storage)}.\\
\end{cases}
\end{align*}
\end{itemize}

We describe the design of the predictive control of PV power generation equipment. We define the vector consisting of the transmission patterns from time~$t=kT_{\rm mini}$ to $t+T_{\rm pred}=k+NT_{\rm mini}$ by 
\begin{align*}
\vb*{{\bf \sigma}}_{{\rm d,pv}}[k]:= 
\begin{bmatrix}
\sigma_{{\rm d,pv}}[k]\\
\sigma_{{\rm d,pv}}[k+1]\\
\vdots \\
\sigma_{{\rm d,pv}}[k+N]\\
\end{bmatrix}.
\end{align*}
We enumerate all~$2^{N}$ possible transmission patterns~$\vb*{\mathbf{\sigma}}_{\rm d,pv}[k]$, 
and compute the value of the evaluation function~$J_{\rm pv}[k]$ corresponding to each transmission pattern. 
Then, the optimal transmission pattern~$\vb*{\mathbf{\sigma}}_{\rm d,pv}^{*}[k]\in\left\{0,1\right\}^{N}$ is expressed as 
\begin{align*}
\vb*{\mathbf{\sigma}}_{\rm d,pv}^{*}[k]
=\underset{\vb*{\mathbf{\sigma}}_{\rm d,pv}[k]\in \left\{0,1\right\}^{N}} {\operatorname{argmin}} \ J_{\rm pv}[k].
\end{align*}
During the time interval from time~$t=kT_{\rm mini}$ to the next switching time~$t+T_{\rm mini}=(k+1)T_{\rm mini}$, 
the control input~$\sigma_{{\rm opt,pv}}(t)$ is given by 
\begin{align*}
\sigma_{{\rm opt,pv}}(t)&=\sigma_{{\rm d,pv}}^{*}(kT_{\rm mini}) , \ t \in [kT_{\rm mini},(k+1)T_{\rm mini}).
\end{align*}

\subsection{Optimal Discharging of Battery  Storage}
\label{sec:battery}

In this subsection, we provide an evaluation function that determines the destination of the power discharged from the battery storage. 
We define the evaluation function~$J_{\rm battery}[k]$ by 
\begin{align*}
J_{\rm battery}[k]:=  w_{\rm batteryfluc}V_{\rm batteryfluc}[k]
+w_{\rm stor,2}B_{\rm stor,2}[k].
\end{align*}
\begin{itemize}
\item
\underline{The first term}: The function 
$V_{\rm batteryfluc}[k]$~[V] correspond to the evaluation of the voltage fluctuation of the underground power distribution line in the time interval~$[t, t+T_{\rm pred})$, 
and is defined in the same way as in the definition~(\ref{vhendou}).
Moreover, $w_{\rm batteryfluc}$ is the weighting coefficient of its evaluation. 
\item
\underline{The second term}: The function
\begin{align*}
B_{\rm stor,2}{[k]}:=  {\sum_{{l}=1}^{N}\int_{(k+l-1)T_{\rm mini}}^{(k+l)T_{\rm mini}}}\zeta_{\rm stor,2}(t) \mathrm{d}t
\end{align*}
corresponds to the evaluation of the amount of electricity in battery storage in the time interval~$[t, t+T_{\rm pred})$, 
where $\zeta_{\rm stor,2}(t)$ is defined  as follows corresponding to the power transmissions: 
\begin{align*}
\zeta_{\rm stor,2}(t):= 
\begin{cases}
\frac{1}{P_{\rm battery}(t)} & \mbox{\begin{tabular}{@{\,}l}(Underground power\\[-1mm]
distribution line\\[-1mm]
or heating cable)\\
\end{tabular}}.\\
0 & \text{(No power transmission)}\\
\end{cases}
\end{align*}
In addition, $w_{\rm stor,2}$ represents the weighting coefficient of the evaluation. 
\end{itemize}


We define the vector consisting of the discharging pattern from $t=kT_{\rm mini}$ to $t=k+T_{\rm pred}=k+NT_{\rm min}$ by 
\begin{align*}
\vb*{\mathbf{\sigma}}_{\rm d,battery}[k]&:= 
\begin{bmatrix}
\vb*{\mathbf{\sigma}}_{{\rm d,battery},0}[k]\\
\vb*{\mathbf{\sigma}}_{{\rm d,battery},1}[k]\\
\vb*{\mathbf{\sigma}}_{{\rm d,battery},2}[k]\\
\end{bmatrix}, \\
\vb*{{\bf \sigma}}_{{\rm d,battery},i}[k]&:= 
\begin{bmatrix}
\sigma_{{\rm d,battery},i}[k]\\
\sigma_{{\rm d,battery},i}[k+1]\\
\vdots \\
\sigma_{{\rm d,battery},i}[k+N]\\
\end{bmatrix}
 \ (i=0, 1, 2).
\end{align*}
We enumerate all $3^{N}$ patterns of $\vb*{\mathbf{\sigma}}_{\rm d,battery}[k]$, 
and perform predictive simulations of the switching control. 
Then, the optimal discharging pattern $\vb*{\mathbf{\sigma}}_{\rm d,battery}^{*}[k]\in\left\{0,1\right\}^{N}$ at $t=kT_{\rm mini}$ is expressed as 
\begin{align*}
\vb*{\mathbf{\sigma}}_{\rm d,battery}^{*}[k]
=\underset{\vb*{\mathbf{\sigma}}_{\rm d,battery}[k]\in \left\{0,1\right\}^{N}} {\operatorname{argmin}} \ J_{\rm battery}[k].
\end{align*}
Then, during the time interval from $t=kT_{\rm mini}$ to the next switching time~$t+T_{\rm mini}=(k+1)T_{\rm mini}$, 
we continue to apply the optimal control input
\begin{align*}
\sigma_{{\rm opt,battery},i}(t)=\sigma_{{\rm d,battery},i}^{*}(kT_{\rm mini}) \ (i=0,1,2),& \\ 
t \in [kT_{\rm mini},(k+1)T_{\rm mini}).&
\end{align*}

\section{Numerical Verification}
\label{sec:nume}

In this section, we verify the effectiveness of the proposed predictive switching control 
via numerical simulations considering two cases of typical power consumption, PV power generation, and weather conditions. 
These simulations suppose the winter season of Toyama Prefecture which is one of the heaviest snowfall areas in Japan. 
The simulation setting is given in Subsection ~\ref{sec:sett}. In addition, the actual time series data for power consumption, PV power generation, and weather conditions used in this simulation are described in Subsection~\ref{sec:data}.
Moreover, the simulation in Subsection~\ref{simyupata-n1} considers the morning of a sunny winter day. Finally, in Subsection~\ref{simyupata-n2}, we deal with a situation in which snowfall occurs, assuming an evening in winter. In particular, in both simulations, a comparison with the results of previous study~\cite{sice} is given to illustrate the contribution of this paper.

\subsection{Simulation Setting}
\label{sec:sett}

In this subsection, we provide the setting of the power distribution system considered in the numerical simulation. 
We consider a power distribution system consisting of the underground power distribution line and the heating cable with length $100$~m. 
The single-line diagram of the system is illustrated in Fig.~\ref{pata-n1}. 
For the underground power distribution line, 
the reference voltage amplitude for p.u. value is set to $v_{1}=\frac{6600}{\sqrt{3}}$~[V]. 
Since we deal with only one of the three phases of the underground power distribution line in this paper, we convert this reference value of line voltage~$\sqrt{3}v_{1}=6600$~[V] to a phase voltage. 
In the heating cable, the reference voltage amplitude for the p.u. value is set to $\frac{1}{a_{1}}v_{1}=200$~[V], 
where $a_{1}=\frac{33}{\sqrt{3}}$ is the transformer ratio. 
The load electrical equipment is installed at the horizontal position $x=25$~[m] from Bifurcation point~1, 
and the PV power generation equipment {is installed} at $x=50$~[m]. 
In addition, the battery storage is connected at $x=75$~[m]. 
{{The connecting position~$x_{\rm load}$ [m] of the load of the heating cable is given by (\ref{hload}). 
We assume that no reactive power is supplied at all position~$x$ of the underground power distribution line and heating cable, i.e. $q_{\rm e}(x,t)=q_{\rm h}(x,t)=0$~[p.u.]. 

\begin{figure}[htb]
\centering
\includegraphics[width=0.95\columnwidth]{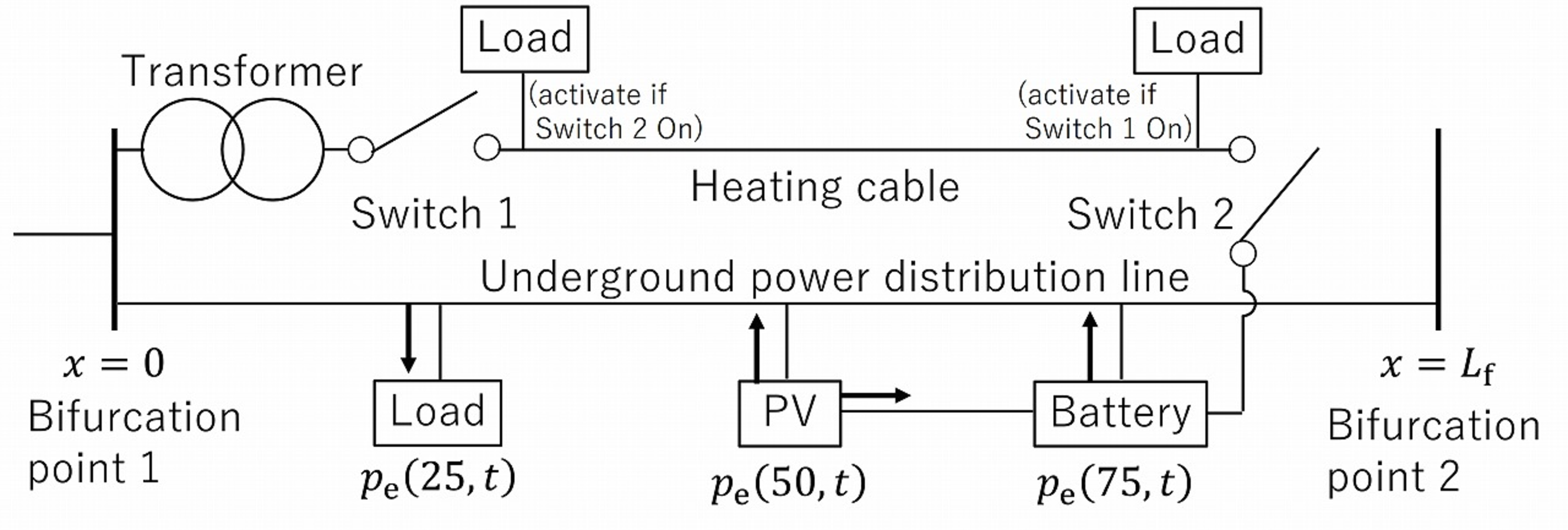}
\caption{{Single-line diagram of power distribution system considered in numerical simulation}}
\label{pata-n1}
\end{figure}

We describe the parameters used in the simulations of this section.
The parameters related to the voltage distribution in the underground power distribution line and heating cable are shown in Tables~\ref{kijyunti} and \ref{p.u.value}, respectively.
The parameters related to the temperature distribution of the heating cable, the thermal diffusion in the underground, and the snow melting on the ground surface are shown in Table~\ref{ondobunpuparame-ta}. 
The parameters related to 
the evaluation function are summarized in Table~\ref{jikanparame-ta}. 
We set the prediction horizon to be $T_{\rm pred}=30$~[min] 
from the viewpoint of the prediction of temporal changes of snow melting against the occurrence of pedestrian falls and traffic congestion due to snow accumulation. 
Moreover, the minimum switching time is set to $T_{\rm mini}=10$~[min], considering the influence of frequent switching on the power distribution system. 


\begin{table}[htb]
\caption{
{Reference values of voltage profile~\cite{odesim}}}
\label{kijyunti}
\centering
\begin{tabular}{@{\,}c@{\,}|@{\,}c@{\,}|@{\,}c@{\,}}
\hline
Parameter & Symbol & Value\\
\hline 
Apparent power & $W_{\rm base}$ & $10$~[kVA]\\ 
Length of underground 
power distribution line 
& $L_{\rm f}$ & $100$~[m]\\
\begin{tabular}{@{\,}l@{\,}}
Voltage phase of underground \\
power distribution line \\ 
\end{tabular} 
& $\theta_{1}$  & $0$~[rad]\\
\begin{tabular}{@{\,}l@{\,}}
Voltage amplitude of underground \\
power distribution line \\  
\end{tabular} 
& $v_{1}$ & $\frac{6600}{\sqrt{3}}$~[V]\\
Voltage amplitude of heating cable 
& $\frac{1}{{a_{\rm 1}}}v_{1}$ & $200$~[V]\\
{Transformer ratio} & ${a_{\rm 1}}$ & $\frac{33}{\sqrt{3}}$\\
Conductance & $g_{\rm base} $ & $918$~[m/$\Omega$]\\
Susceptance & $b_{\rm base}$ & $918$~[m/$\Omega$]\\
\hline
\end{tabular}
\end{table}

\begin{table}[htb]
\caption{{p.u. values of voltage profile~\cite{odesim}}}
\label{p.u.value}
\centering
\begin{tabular}{@{\,}c@{\,}|@{\,}c@{\,}|@{\,}c@{\,}}
\hline
Parameter & Symbol & p.u. value\\
\hline 
\underline{Underground power distribution line} & & \\
Length & $L_{\rm f}$ &$1$\\
Voltage amplitude & $v_{1}$ & $1$\\
Conductance  & $g_{\rm e}$ & $1$\\
Susceptance & $b_{\rm e}$ & $1$\\ \hline 
\underline{Heating cable} & & \\
Length & $L_{\rm f}$ &$1$\\
Conductance & $g_{\rm h} $ & $0.5$\\
Susceptance & $b_{\rm h}$ & $0.5$\\
Active power of load & $p_{\rm h}(x_{\rm load},t)$ & $-10$\\
\hline
\end{tabular}
\end{table}


\begin{table}[htb]
\caption{
Parameters related to heating cable, thermal diffusion, and snow melting~\cite{netusyusi}}
\label{ondobunpuparame-ta}
\centering
\begin{tabular}{@{\,}c@{\,}|@{\,}c@{\,}|@{\,}c@{\,}}
\hline
Parameter & Symbol & Value\\
\hline 
\underline{Heating cable}
& & \\
Radiative cooling & $q_{\rm r}$ & $0$~[W/cm]\\
{\begin{tabular}{@{\,}l@{\,}}
 Contact heat transfer coefficient  \\
 per unit length \\
 between cable surface and soil\\ 
   \end{tabular}} & $\gamma_{\rm cable}$ & $1.04$~[W/m$\cdot$K]\\
 Heat capacity & $C_{\rm cable}$ & $18$~[J/cm$\cdot{}^\circ$C]\\
 Thermal conductivity of soil & $\lambda_{\rm soil}$ & $0.5$~[W/m$\cdot$K]\\
\begin{tabular}{l}
Heat transfer coefficient \\
between heating cable and soil\\ 
\end{tabular} 
 & {$\beta_{\rm cable}$} & {$300$~[W/m${}^{2}\cdot$K]}\\
\hline
\underline{Thermal diffusion}
& & \\ 
Burial depth of heating cable & $D_{\rm f}$ & $10$~[cm]\\
 Thermal diffusivity of soil & $\alpha_{\rm soil}$ & $0.008$~[m${}^2$/s]\\
\begin{tabular}{l}
Heat transfer coefficient \\
between ground surface and snow\\ 
 \end{tabular}
& {$\beta_{\rm ground}$} & {$88$~[W/m${}^{2}\cdot$K]}\\ \hline
\underline{Snow melting}
& & \\ 
Snow temperature & $\delta_{\rm snow}$ & $0$~[${}^\circ$C]\\ 
 Initial condition for snow volume & $h_{\rm snow}(x,0)$ & $30$~[mm]\\
 Unit conversion factor & $a_{\rm snow}$ & $1.792\times 10^{-4}$\\
 Density of snow volume & $d_{\rm snow}$ & $0.06$~[g/cm${}^{3}$]\\
\hline
\end{tabular}
\end{table}


\begin{table}[htb]
\caption{{Parameters related to evaluation function}}
\label{jikanparame-ta}
\centering
\begin{tabular}{@{\,}c@{\,}|@{\,}c@{\,}|@{\,}c@{\,}}
\hline
Parameter & Symbol & Value\\
\hline 
Minimum switching time & $T_{\rm mini}$ & $10$~[min]\\
Predictive horizon & $T_{\rm pred}$ & $30$~[min]\\ \hline
\underline{Weighting coefficients (Heating cable)} & & \\ 
Distribution loss & $w_{\rm loss}$ & $4\times10^2$\\
Voltage fluctuation & $w_{\rm fluc}$ & $1\times10^7$\\
Snow volume & $w_{\rm snow}$ & {$1.2\times10^5$}\\
Cost & $w_{\rm cost}$ & {$8\times10^5$}\\ \hline
\underline{Weighting coefficients (PV)} & & \\ 
Voltage fluctuation & $w_{\rm pvfluc}$ & {$1$}\\ 
Amount of power of battery storage & $w_{\rm stor,1}$ & {$1\times10^{-3}$}\\ \hline
\underline{Weighting coefficients (Battery storage)} & & \\ 
Voltage fluctuation & $w_{\rm batteryfluc}$ & {$1$}\\
Amount of power of battery storage & $w_{\rm stor,2}$ & {$1\times10^{-3}$}\\
\hline
\end{tabular}
\end{table}
\subsection{Time Series Data for Simulation}
\label{sec:data}

In this subsection, we describe the time series data and processing of actual time series data on electricity and weather.

\subsubsection{Power consumption and PV power generation}
\label{app:data_power}

The time series data on the power consumption of the residential loads are provided by Nagoya University Open Data for EMS Evaluation~\cite{phouse} for ten residential houses. 
As we consider only single phase of the three-phase underground power distribution line, {{$\frac{1}{3}$}} of the time series data for $10$~houses is used as the active power~$p_{\rm e}(x,t)$ of the residential load. 

We use time series data of PV power generation provided by Hokuriku Electric Power Transmission and Distribution Company~\cite{pvdata}. 
We also use the value that is $\frac{1}{1000}$ of the data in order to fit our data to the scale of the underground power distribution line considered in this paper. 
Furthermore, as we consider only one phase among the three phases, 
we use {{$\frac{1}{3}$}} of the time series data that fits the scale of the distribution system considered in this paper as the effective power~$p_{\rm e}(x,t)$ of the PV power generation equipment. 
On the other hand, when the {{active power}} is transmitted from the PV power generation equipment to the battery storage, we suppose that all the {{active power}} is  transmitted. 
For this reason, we do not modify the scale of the data which is fit to the scale considered in this paper. 

The power supplied by the PV power generation 
is considered to be more accurately predicted based on the day's weather forecast, compared to the power consumption of the load. 
Thus, we use the time series data for PV power generation as predictive data. 
In general, it is not easy to predict these data exactly.
For this reason, 
the predicted value of the data at the predictive horizon continues to use the value at the current time as a constant value in the computation of the evaluation function.

\subsubsection{Weather}

As the time series data of solar radiation, we use the average data of the radiation from 1981 to 2009 provided by NEDO's database viewing system~\cite{sundata}. 
Since this data is time series data at 1~h intervals, it is linearly interpolated as data at 30~sec intervals.
Moreover, we choose the time series data observed in Toyama City. 
The data is provided by the Japan Meteorological Agency~\cite{temp} for atmospheric temperature and wind speed. 
Since this data is given as time series data every 10~min, linear interpolation is performed to obtain data every 30~sec.

\subsection{Case~1 (Morning without snowfall)}
\label{simyupata-n1}

\subsubsection{Setting and time series data}

In this subsection, 
we consider a numerical simulation for a sunny winter morning with no snowfall from A.M.~6:00 to 8:00.
This implies that that A.M.~6:00 and 8:00 correspond to $t=0$~[min] and $t=120$~[min], respectively. 
For the time series data used in this case, we use January 19, 2016 for the residential load~\cite{phouse}, January 19, 2020 for the PV power generation~\cite{pvdata}. 
Fig.~\ref{phouseppv1} shows the temporal variation of the power consumption of residential loads and PV power generation. 
We also use the time series data for atmospheric temperature~\cite{temp}, wind speed~\cite{temp}, and the average data from January 19, 1981 to 2009 for solar radiation~\cite{sundata}. 
The time series data on these quantities are shown in Fig.~\ref{jitude-ta1}.

\begin{figure}[htb]
\begin{center}
\subfigure[Power consumption of residential loads]{
\includegraphics[width=40mm]
{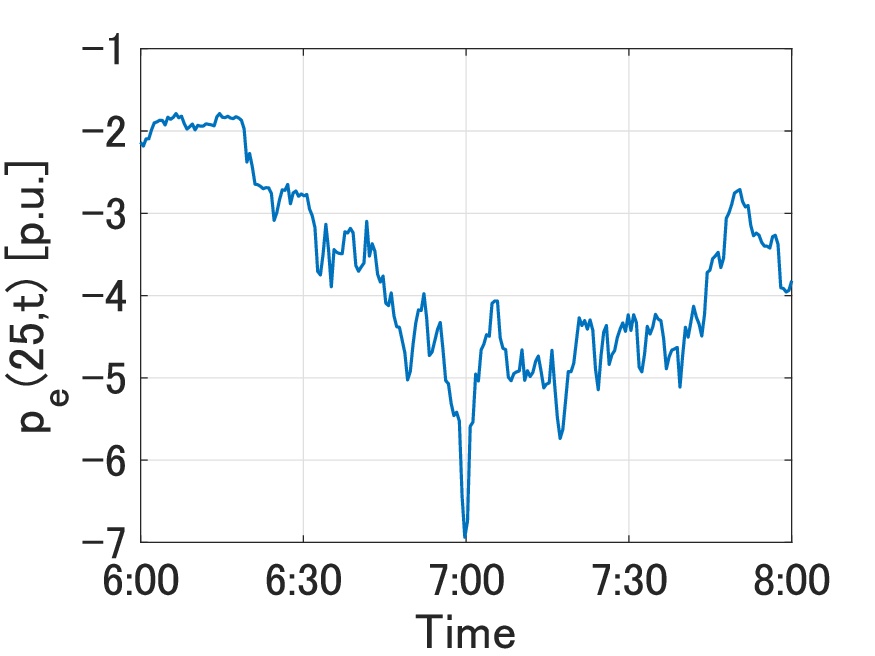}
}
\subfigure[PV power generation]{
\includegraphics[width=40mm]
{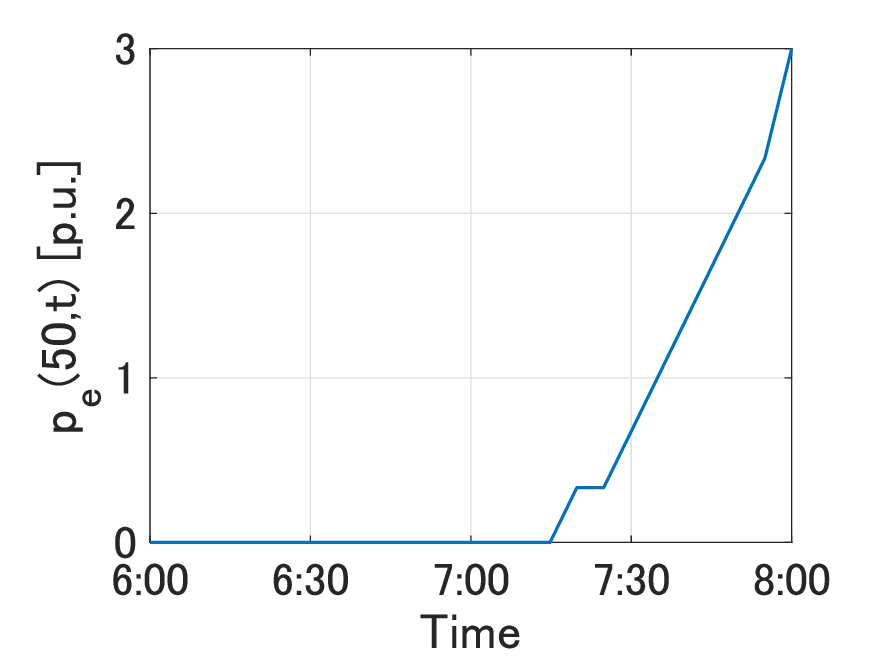}
}
\end{center}
\caption{
Temporal variation of active powers~(Case~1)}
\label{phouseppv1}
\end{figure}


\begin{figure}[htb]
\begin{center}
\subfigure[Solar radiation]{
\includegraphics[width=40mm]
{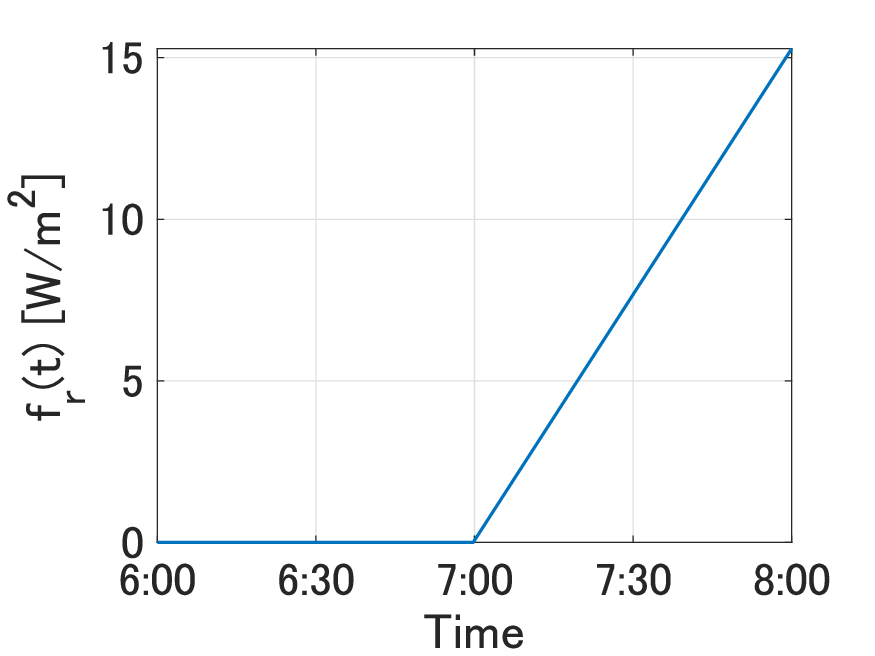}
}
\subfigure[Snowfall]{
\includegraphics[width=40mm]
{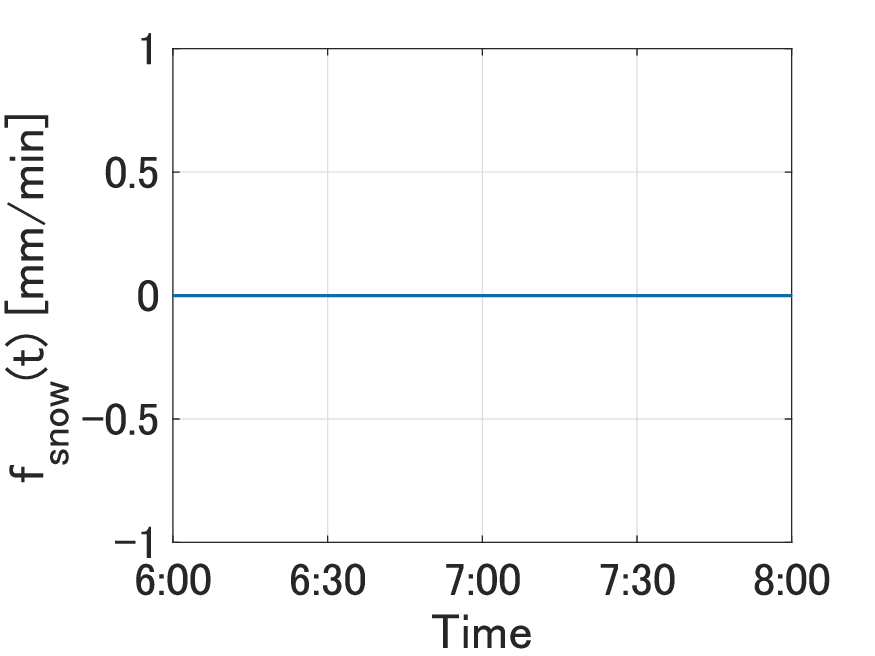}
}\\
\subfigure[Atmospheric temperature]{
\includegraphics[width=40mm]
{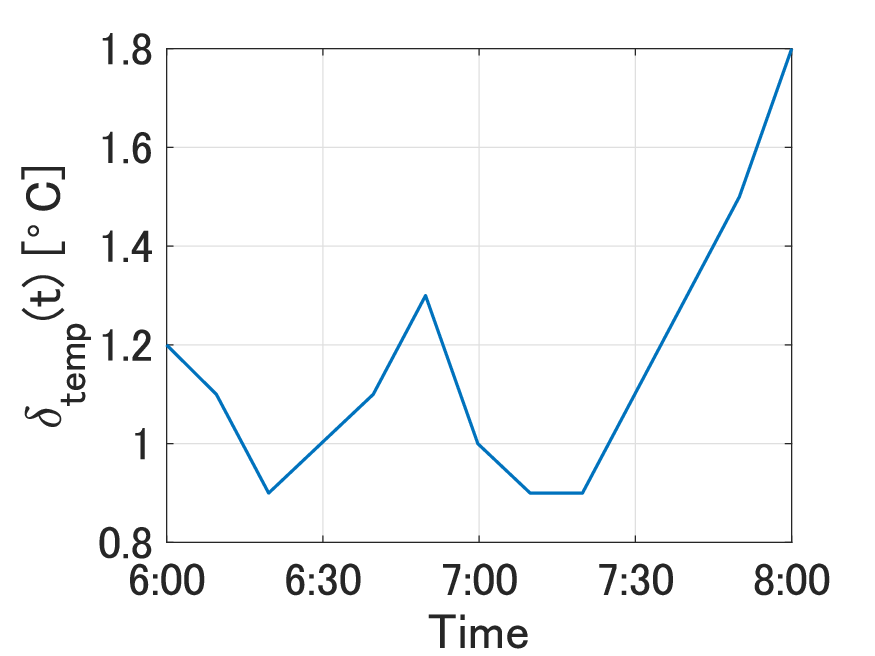}
}
\subfigure[Wind speed]{
\includegraphics[width=40mm]
{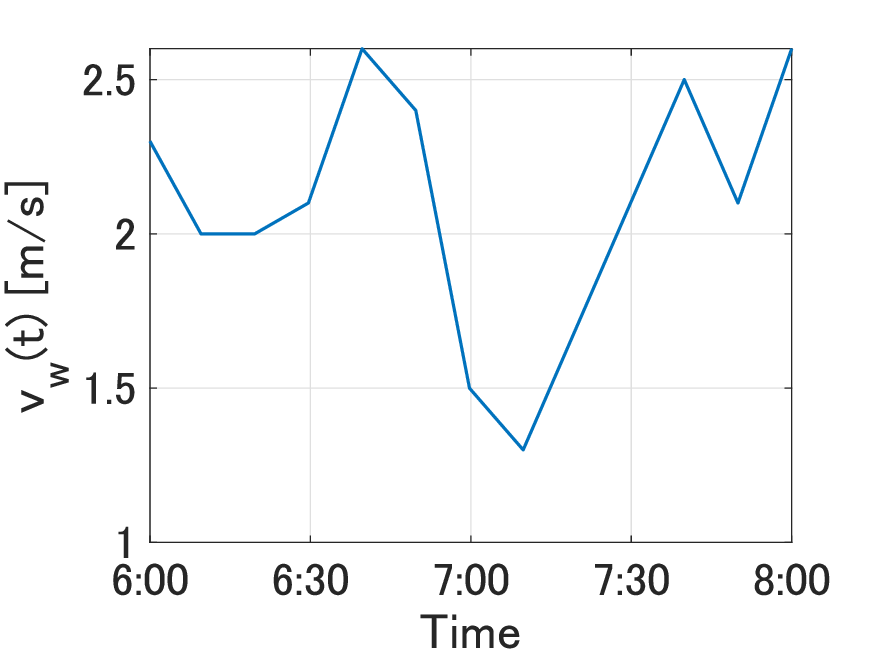}
}
\end{center}
\caption{
Temporal variation of variables related to weather~(Case~1)}
\label{jitude-ta1}
\end{figure}

\subsubsection{Simulation result}
\label{sec:resu1}

\subsubsection*{(1) Temporal variation of variables}

We show the simulation results for Case~1 in Fig.~\ref{simkekka1}. 
In Fig.~\ref{simkekka1} (a), the voltage amplitude~$v_{\rm e}(x;t)$ [V] of the underground power distribution line varies as power consumption of the residential load and heating cable, and power supply from PV {{power generation equipment}}. 
The amplitude is reduced due to the power consumption of the residential load during A.M.~6:50 to 7:00. 
On the other hand, the amplitude increases with the PV power generation between A.M.~7:50 and 8:00. 
Fig.~\ref{simkekka1} (b) shows that the voltage amplitude~$v_{\rm h}(x;t)$ [V] of the heating cable varies greatly as the switches change. 
When Switch~1 is turned {\tt On}, the voltage amplitude decreases from $x=0$~[m] to $x=100$~[m].
On the other hand, the amplitude decreases from $x=100$~[m] to $x=0$~[m] 
during Switch~2 is turned {\tt On}. 

The temperature~$\delta_{\rm surf}(t;x)$ [${}^{\circ}$C] of the cable surface is shown in Fig.~\ref{simkekka1} {(c)}. 
The temperature increases because Switches~1 or 2 are turned {\tt On} {from A.M.~6:00 to 7:30}, and decreases during A.M.~7:30 to 8:00 when both switches are turned {\tt Off}.
When Switch~1 is turned {\tt On}, the temperature at $x=0$~[m] is higher than at $x=100$~[m].
On the other hand, if Switch~2 is turned {\tt On}, the temperature at $x=100$~[m] increases compared to the temperature at $x=0$~[m]. 
From Fig.~\ref{simkekka1} (d), the snow volume~$h_{\rm snow}(t;x)$~[mm] decreases with time {{over the intermediate positions of the heating cable}}. 
{{Moreover, this figure shows that there is a local increase in the snow volume at $x=0$~[m] and $x=100$~[m]. 
The reason for this phenomena can be explained as follows. 
Due to the boundary conditions given in Subsection~\ref{sec:boun}, 
if Switch~1 and Switch~2 are selected, no current flows at $x=100$~[m] and $x=0$~[m], respectively. 
For this reason, we observe less snow melting due to the increase in the surface temperature of the cable because there is no power distribution loss at these positions. 
An improvement of this point remains as our future work. 
Please refer to the second paragraph (iv) of Section~\ref{sec:conc} for specific details.}}

\begin{figure}[htb]
\begin{center}
\subfigure[Voltage amplitude of underground power distribution line]{
\includegraphics[width=40mm]
{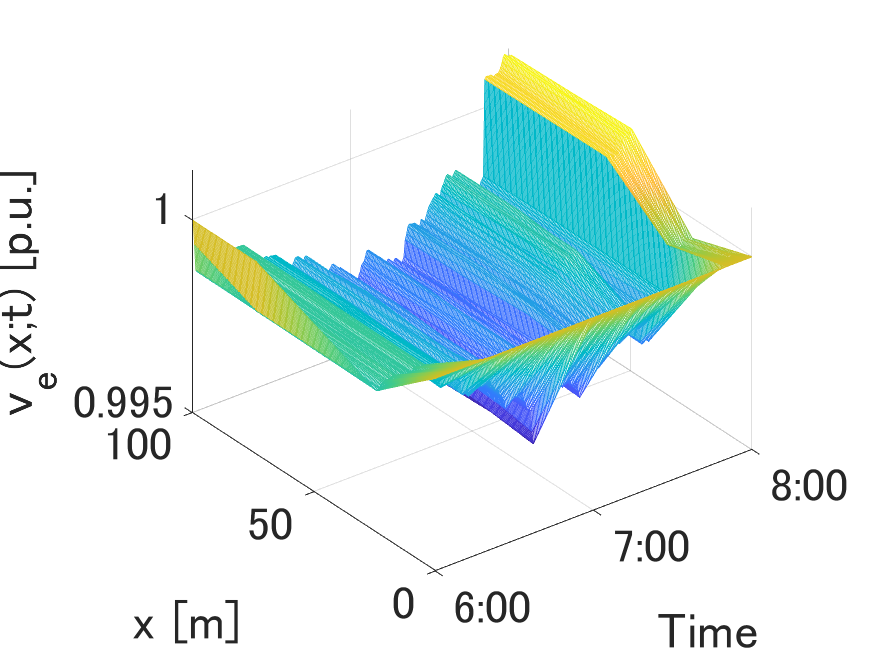}
}
\subfigure[Voltage amplitude of heating cable]{
\includegraphics[width=40mm]
{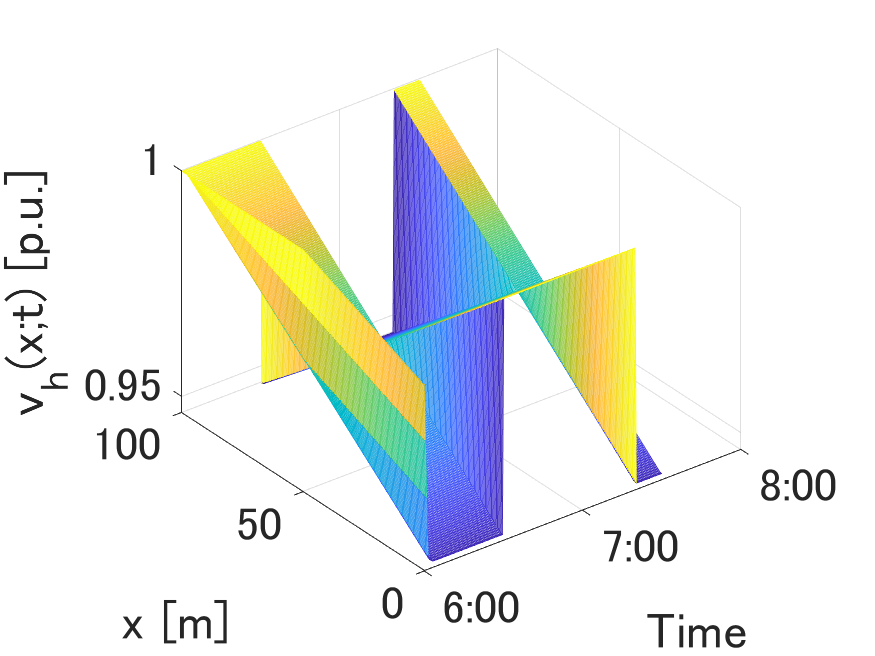}
}\\
\subfigure[Temperature of cable surface]{
\includegraphics[width=40mm]
{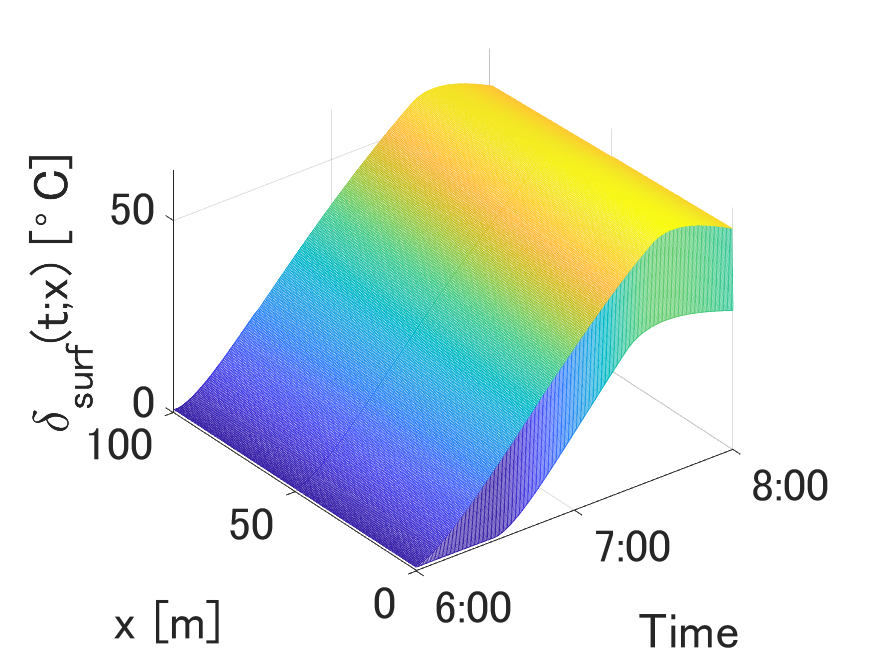}
}
\subfigure[Snow volume]{
\includegraphics[width=40mm]
{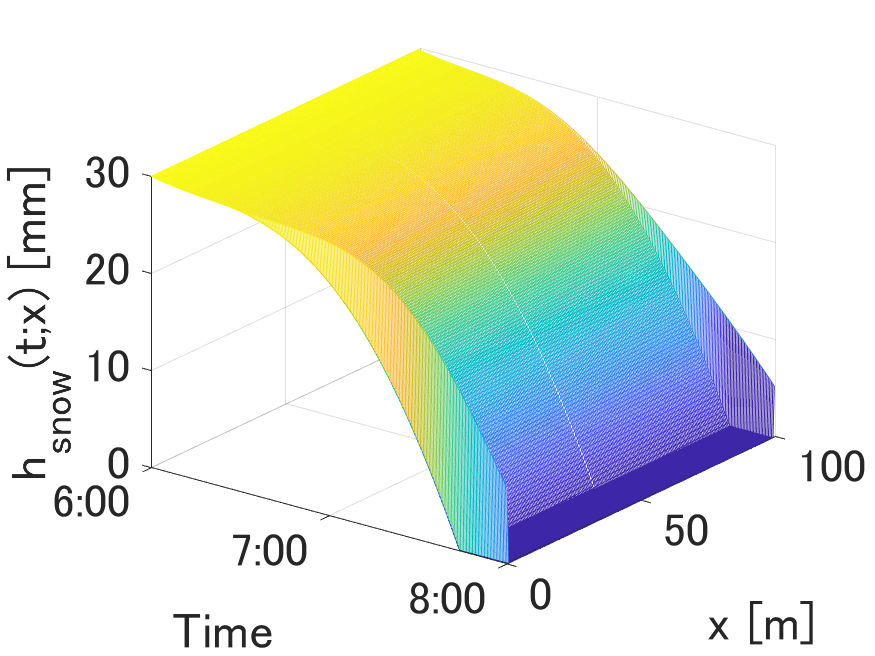}
}
\end{center}
\caption{
Spatial and temporal variation of the variables~(Case~1)}
\label{simkekka1}
\end{figure}

\subsubsection*{(2) Temporal variation of switching and discharging}

We show the simulation results for switching, and the PV power generation equipment, battery storage in Fig.~\ref{simkekka1switch}. 

Fig.~\ref{simkekka1switch} (a) illustrates the temporal variation of the switching control input for Case~1 by showing the index~$i$ which satisfies $\sigma_{i}(t)=1$, the switch that turns {\tt On}. 
From A.M.~6:00 to A.M.~6:30, 
Switch~2, which is powered by a battery storage, is turned to {\tt On}. 
It is assumed that the owners of the heating cable and battery storage are the same, 
and Switch~2 is selected because it is less expensive than Switch~1. 
From A.M.~6:30 to 7:20, Switch~1 is set to {\tt On} to melt snow, 
because the selection of  Switch~2 would violate the constraint on the reserve capacity of the battery storage. From A.M.~7:20 to 7:30, Switch~2 is used for snow melting because the PV power generation system has supplied power to the battery storage. 
From A.M.~7:30 a.m. to 8:00, the switch is set to {\tt Off} because the snow melting is almost completed. 

Next, we show the temporal variation of the control input~$\sigma_{{\rm pv}}(t)$ of the PV power generation equipment in Fig.~\ref{simkekka1switch} (b). 
It can be confirmed that the {{active power}} is transmitted to the battery storage from A.M.~6:00  to 7:20 and from A.M.~7:40 to 7:50, and to the underground power distribution line from A.M.~7:20 to 7:40 and from A.M.~7:50 to 8:00. 

Finally, we show the temporal variation of the index~$i$ satisfying $\sigma_{{\rm battery},i}(t)=1$, i.e. the destination of the {{active power}} of the battery storage 
in Fig.~\ref{simkekka1switch}~(c). 
When Switch~2 is turned {\tt On}, the battery storage always transmits the {{active power}} to the heating cable. In addition, from A.M.~6:30 to 7:20 and from A.M.~7:30 to 8:00, the {{active power}} is stored to the battery without being transmitted anywhere. 
We also present the temporal variation of the amount of electricity in the battery storage in Fig.~\ref{simkekka1switch}~(d). 
From A.M.~6:00 a.m. to 6:30 and from A.M.~7:20 a.m. to 7:30, the amount of {{the active power}} decreases with time because the battery storage transmits power to the heating cables. 
From A.M.~7:10 to 7:20 a.m. and from A.M.~7:40 to 7:50 a.m., the amount of {{the active power}} increases because the PV power generation equipment transmits power while the battery storage is not transmitting {{the active power}} anywhere. 

\begin{figure}[htb]
\begin{center}
\subfigure[Temporal variation of switching control input]{
\includegraphics[width=40mm]
{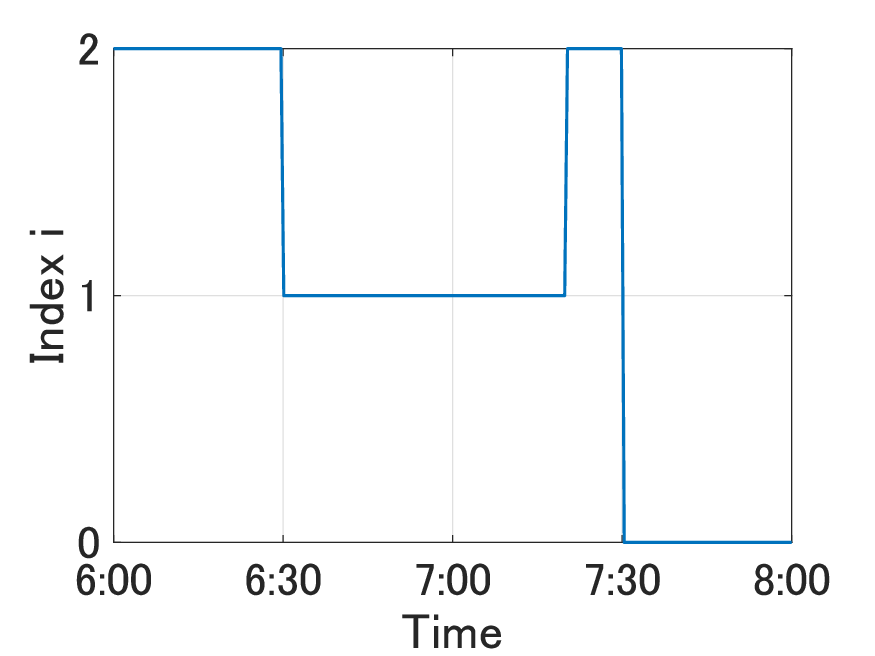}
}
\subfigure[Temporal variation of the destination of PV power]{
\includegraphics[width=40mm]
{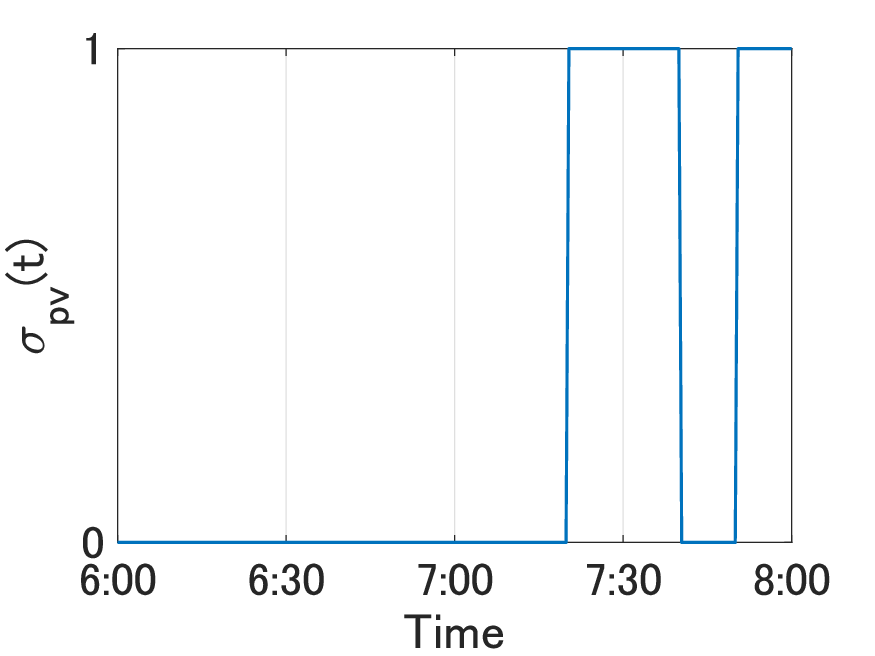}
}\\
\subfigure[Temporal variation of the destination of battery power]{
\includegraphics[width=40mm]
{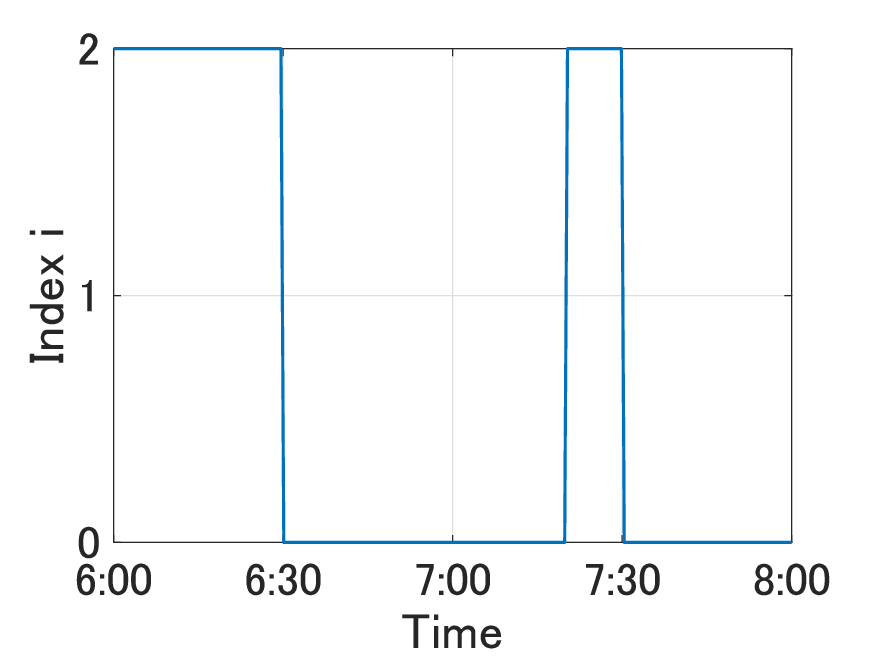}
}
\subfigure[Temporal variation of the amount of electricity in battery]{
\includegraphics[width=40mm]
{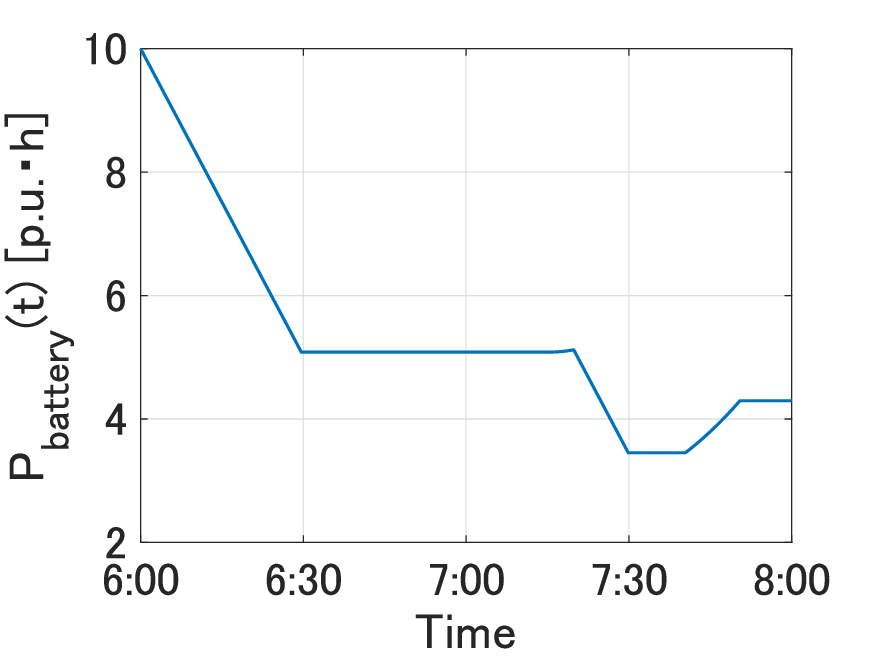}
}
\end{center}
\caption{
Temporal variation with respect to switches, PV, and battery storage~(Case~1)}
\label{simkekka1switch}
\end{figure}

\subsubsection*{(3) Comparison with and without battery storage}

In the following, 
we compare the cost and voltage variations without and with battery storage. 
Note that the former and latter simulations correspond to the result of the switching control due to the authors' previous study~\cite{sice} and this paper, respectively. 

We show the temporal variation of the switches for the cases without and with the battery storage in Fig.~\ref{hikaku1}~(a) and Fig.~\ref{hikaku1}~(b), respectively. 
In Fig.~\ref{hikaku1}~(a), $15$~[p.u.$\cdot$h] of electricity is purchased because Switch~1 is selected from A.M.~6:00 am to 7:30. 
On the other hand, in Fig.~\ref{hikaku1}~(b), Switch~1 is selected from A.M.~6:30 to 7:20, and thus $8.3$~[p.u.$\cdot$h] of electricity is purchased. 
Thus, with the introduction of the battery storage, we can confirm that the cost of using the heating cables is reduced. 

Figs.~\ref{hikaku1}~(c) and (d) show the temporal variation of the voltage fluctuation of the underground power distribution line without and with the battery storage, respectively. 
Comparing the temporal variations for each case, there is no significant difference between the two cases. 
In Fig.~\ref{hikaku1}~(c), if no battery storage is installed, 
the PV power generation equipment transmits all the {{active power}} to the underground power distribution line. 
On the other hand, in Fig.~\ref{hikaku1}~(d), the 
equipment transmits power to the underground power distribution line and the battery storage. 
This shows the fact that the installation of the storage efficiently suppresses the voltage fluctuations while recharging the power, 
which implies the efficiency of the switching control of this paper comparing with \cite{sice}. 

{{Finally, Fig.~\ref{hikaku1} (e) shows the voltage fluctuations with and without a battery storage. 
We can see that the installation of the battery storage suppresses the voltage fluctuation of the underground power distribution line during most of the time periods. However, this installation does not necessarily suppress the fluctuations during A.M.~7:10-7:20 and A.M.~7:40-7:50. 
In the case with the battery storage, during these time periods, the PV power generation equipment does not transmit active power to the underground power distribution line but charges the battery storage. 
This shows that the effective use of surplus power, which does not purchase commercial power, is given priority over the suppression of voltage fluctuations to the distribution line. 
Another reason is that, with the installation of the battery storage, 
the active power is supplied from the battery storage to the heating cables during A.M.~7:20-7:30. 
This shows that the snow melting is given priority over voltage fluctuation suppression. 
It is thought that the switching selection during this time period has affected the voltage fluctuations during the time period under consideration.}} 

\begin{figure}[htb]
\begin{center}
\subfigure[Temporal variation of switching control input~(without battery storage)]{
\includegraphics[width=40mm]
{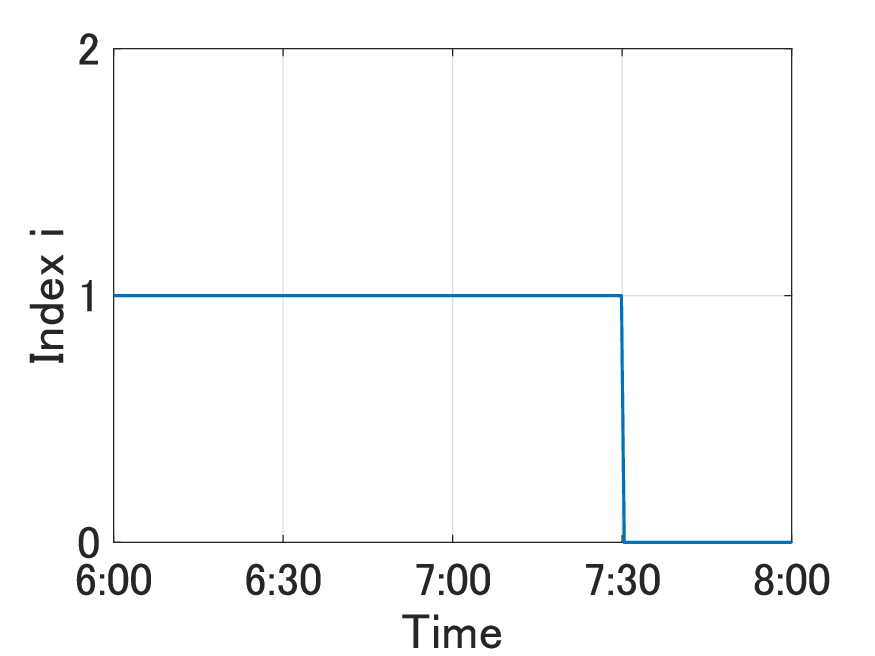}
}
\subfigure[Temporal variation of switching control input~(with battery storage)]{
\includegraphics[width=40mm]
{switch1.eps}
}\\
\subfigure[Voltage fluctuation~(without battery storage)]{
\includegraphics[width=40mm]
{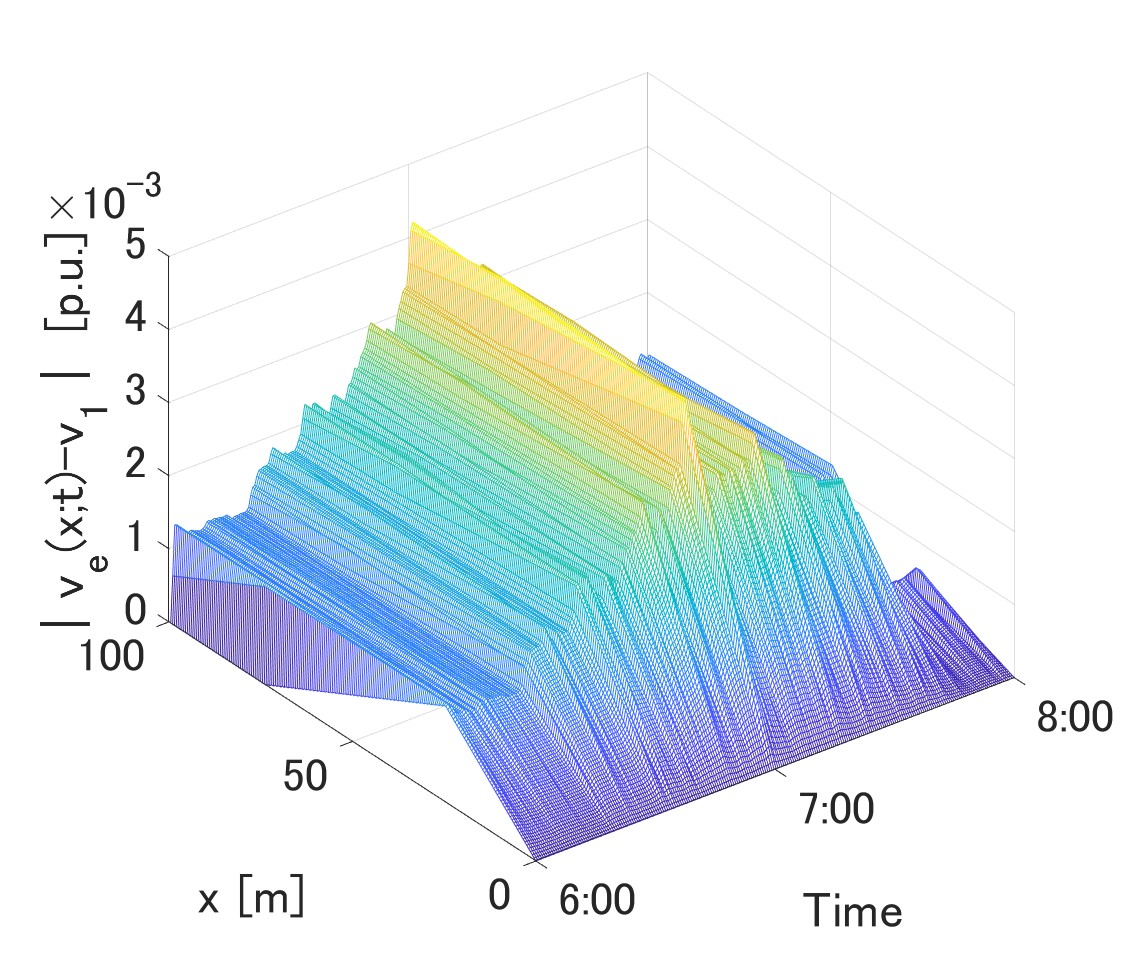}
}
\subfigure[Voltage fluctuation~(with battery storage)]{
\includegraphics[width=40mm]
{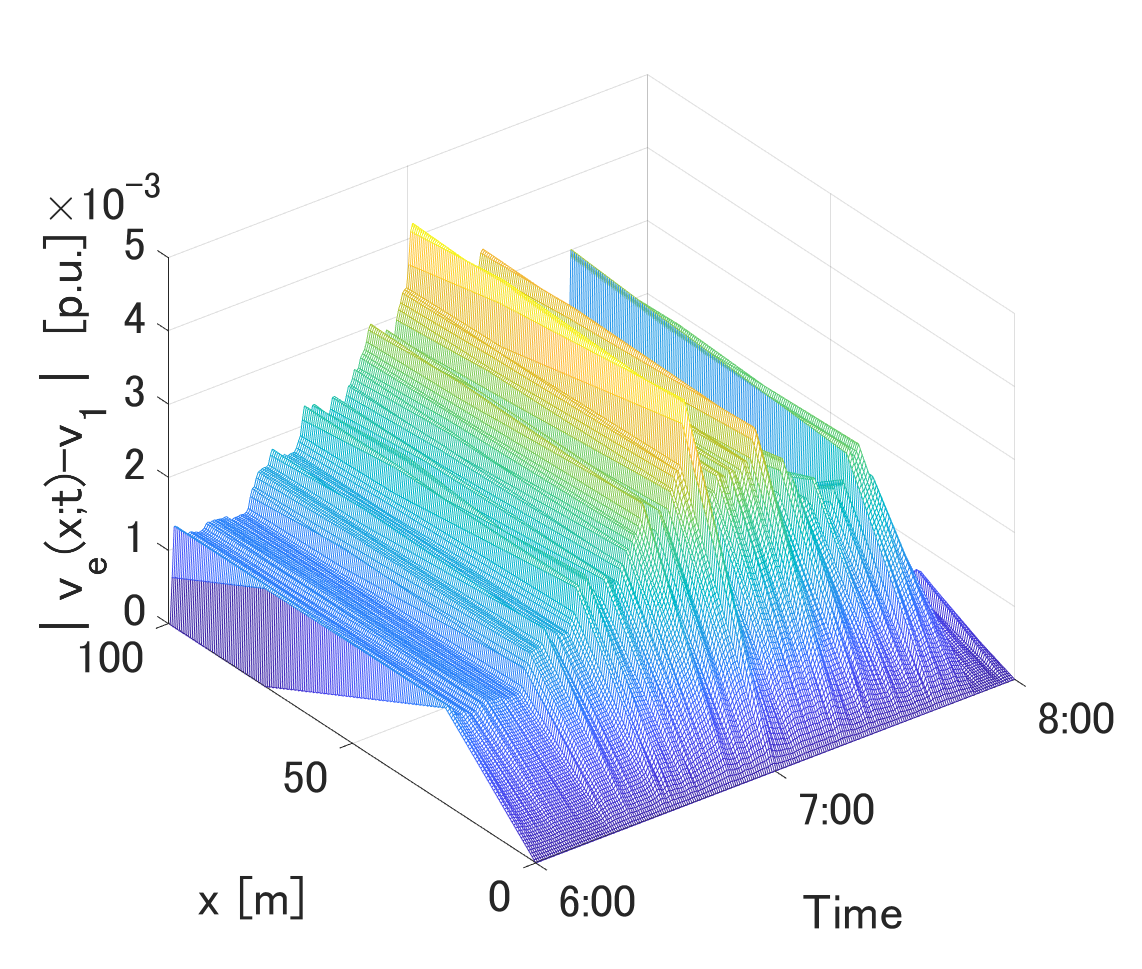}
}\\
\subfigure[{{Difference of voltage fluctuations}}]{
\includegraphics[width=40mm]
{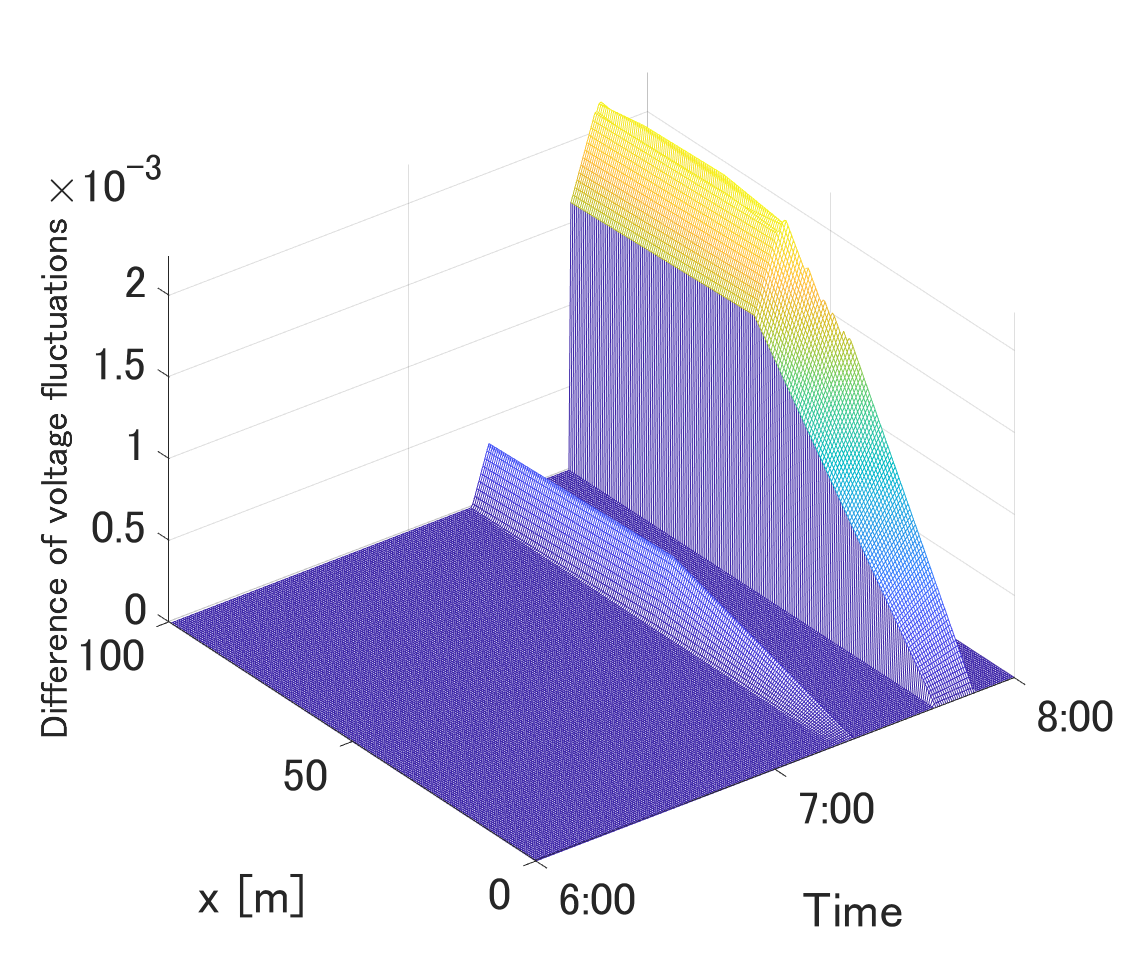}
}
\end{center}
\caption{
Comparison of switching control input and voltage fluctuation with respect to battery storage~(Case~1)}
\label{hikaku1}
\end{figure}

{{For the numerical simulations run with the setting of Case~1, 
the computation times for Section~4.3 of the previous study~[9], where no storage batteries are installed, and this paper are $323.7$~sec and $479.3$~sec, respectively. 
Since this paper also performs the switching of the PV power generation and battery storage, 
the computation time is slightly higher than in [9]. 
However, even for the computation of this paper, which required a large number of computation time, the computation time required for a single switching is $479.3 \times \frac{10}{120} \simeq 39.9$~[sec]. 
The computation is completed in a sufficiently short time compared to the minimum switching time~$T_{\rm mini}=10$~[min]. Therefore, the switching predictive control in this paper is suitable for implementation in real time.}}

\subsection{Case~2 (Evening with snowfall)}
\label{simyupata-n2}
\subsubsection{Setting and time series data}

In this subsection, we consider a numerical simulation for a winter evening from P.M.~4:00 to 6:00 with snowfall starting from P.M.~5:00. 
This implies that P.M.~4:00 and 6:00 correspond to $t=0$~[min] and $t=120$~[min], respectively. 
We use the time series data of the residential load~\cite{phouse} from February 7, 2016, 
and of the PV power generation~\cite{pvdata} from February 7, 2020. 
The temporal variation of the power consumption of residential load and PV power generation is shown in Fig.~\ref{phouseppv2}. 
The time series data for atmospheric temperature and wind speed are from February 7, 2020~\cite{temp}. 
Moreover, we choose the average solar radiation data from February 7, 1981 to 2009~\cite{sundata}. 
The snowfall time series data gives a hypothetical snowfall of $12$~mm per $1$~hour assumed for Toyama Prefecture. 
The time series data of the physical quantities related to weather are shown in Fig.~\ref{jitude-ta2}.

\begin{figure}[htb]
\begin{center}
\subfigure[Power consumption of residential loads]{
\includegraphics[width=40mm]
{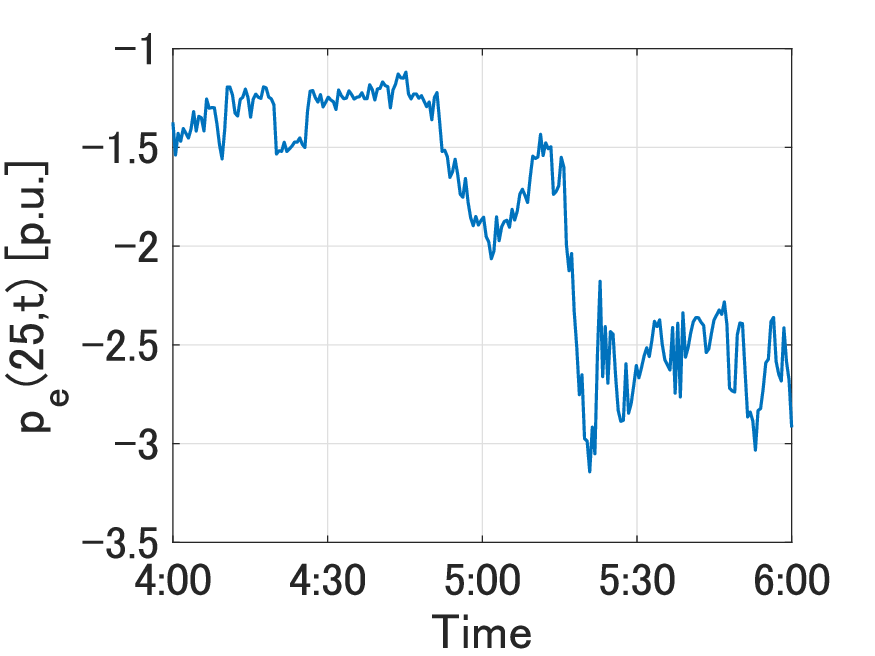}
}
\subfigure[PV power generation]{
\includegraphics[width=40mm]
{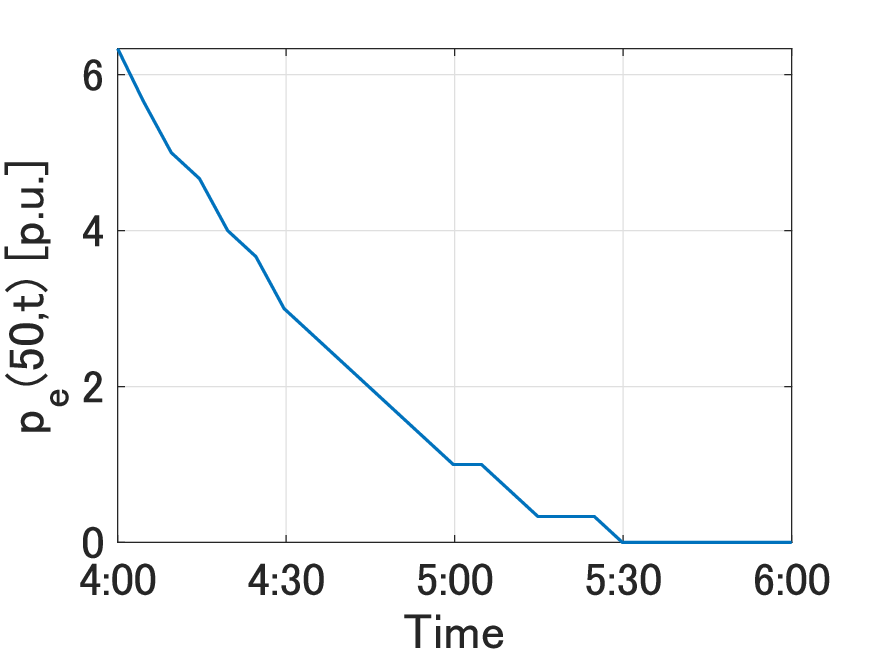}
}
\end{center}
\caption{
Temporal variation of active powers (Case~2)}
\label{phouseppv2}
\end{figure}



\begin{figure}[htb]
\begin{center}
\subfigure[Solar radiation]{
\includegraphics[width=40mm]
{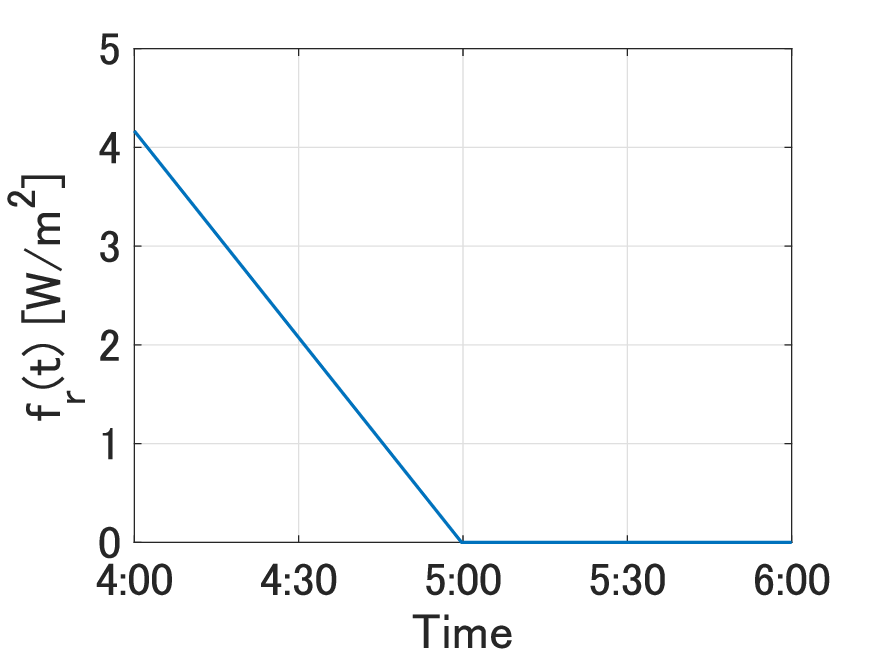}
}
\subfigure[Snowfall]{
\includegraphics[width=40mm]
{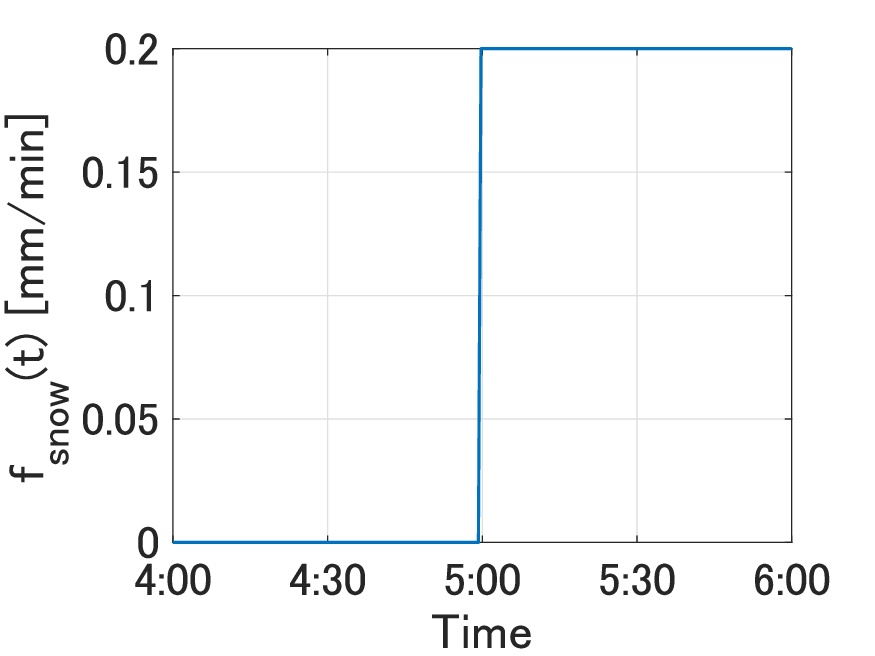}
}\\
\subfigure[Atmospheric temperature]{
\includegraphics[width=40mm]
{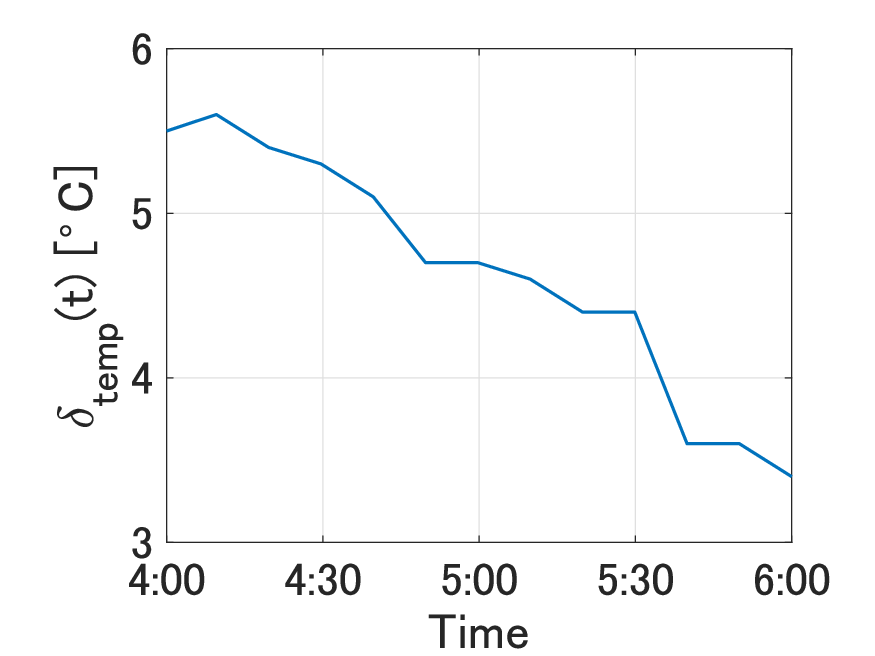}
}
\subfigure[Wind speed]{
\includegraphics[width=40mm]
{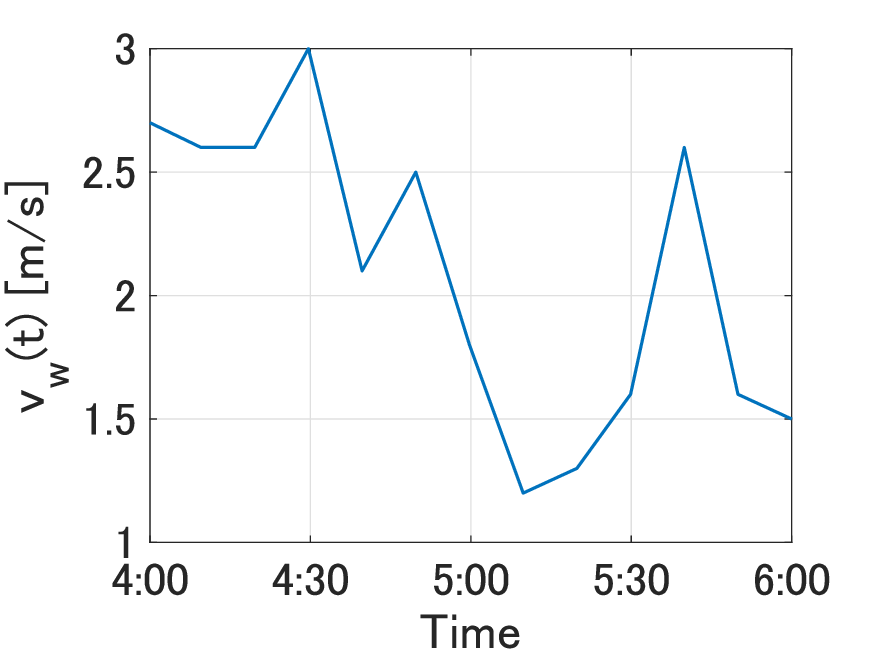}
}
\end{center}
\caption{
Temporal variation of variables related to weather (Case~2)}
\label{jitude-ta2}
\end{figure}

\subsubsection{Simulation result}
\label{sec:resu2}

\subsubsection*{(1) Temporal variation of variables}

The simulation results for Case~2 are shown in Fig.~\ref{simkekka2_mscs}. 
We show the temporal and spatial variations of the voltage amplitude of the underground power distribution line in Fig.~\ref{simkekka2_mscs}~(a). 
In this figure, the voltage fluctuation is suppressed by the power supplied by the PV power generation equipment against the increase of power consumption by the residential load from P.M.~5:00 to 5:30. 
In Fig.~\ref{simkekka2_mscs}~(b), the voltage amplitude of the heating cable decreases from $x=100$~[m] to $x=0$~[m] because Switch~2 is set to {\tt On} at many time intervals. 

We show the temporal and spatial variations of the outer surface temperature ~$\delta_{\rm surf}(t;x)$~[${}^{\circ}$C] of the heating cable in Fig.~\ref{simkekka2_mscs}~(c). 
{{The temperature at $x=100$~[m]}} is higher than that at $x=0$~[m], because Switch~2 is set to {\tt On} during many time periods. 
Moreover, in Fig.~\ref{simkekka2_mscs}~(d), 
The occurrence of snowfall from P.M.~5:00 has changed the decrease of the snow accumulation~$h_{\rm snow}(t;x)$~[mm]. 
{{Fig.~\ref{simkekka2_mscs}~(d) shows a local increase in the snow volume at $x=0$~[m], analogous to the previous section (Case~1). 
The cause of this increase is same as the previous section. 
In particular, Switch~2 is selected from P.M.~4:00 to P.M.~5:30, 
a snow melting due to residual heat cannot be expected at $x=0$ during this period. 
Moreover, the two additional facts can be observed. 
One fact is that, since the sunset was approaching after P.M.~4:00, 
the effect of snow melting due to solar radiation was negligible. 
Another fact is that the snowfall begins at around P.M.~5:00, resulted in a local increase in snowfall. 
An idea to resolve this difficulty will be discussed in the second paragraph (iv) of Section \ref{sec:conc}.}} 

\begin{figure}[htb]
\begin{center}
\subfigure[Voltage amplitude of underground power distribution line]{
\includegraphics[width=40mm]
{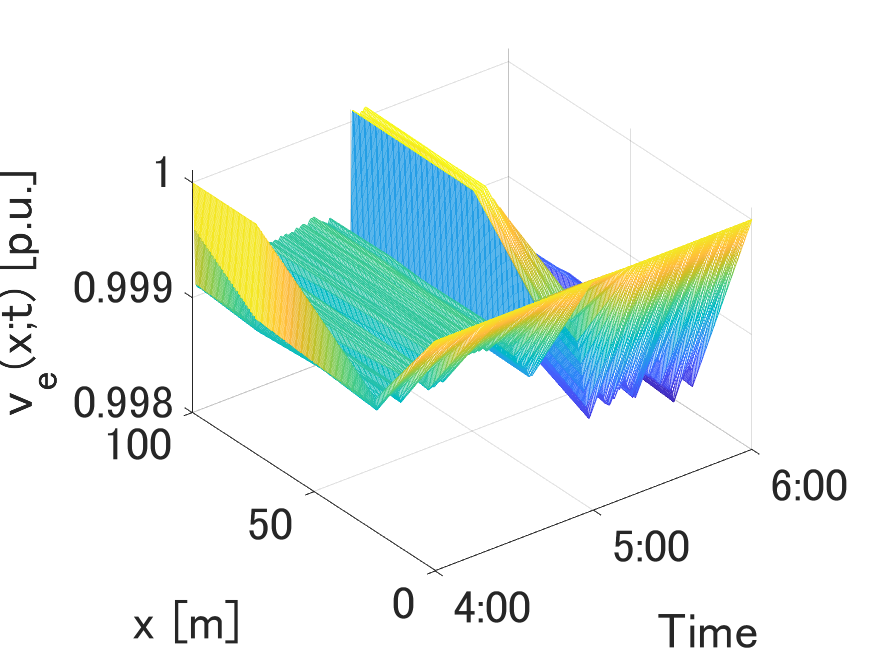}
}
\subfigure[Voltage amplitude of heating cable]{
\includegraphics[width=40mm]
{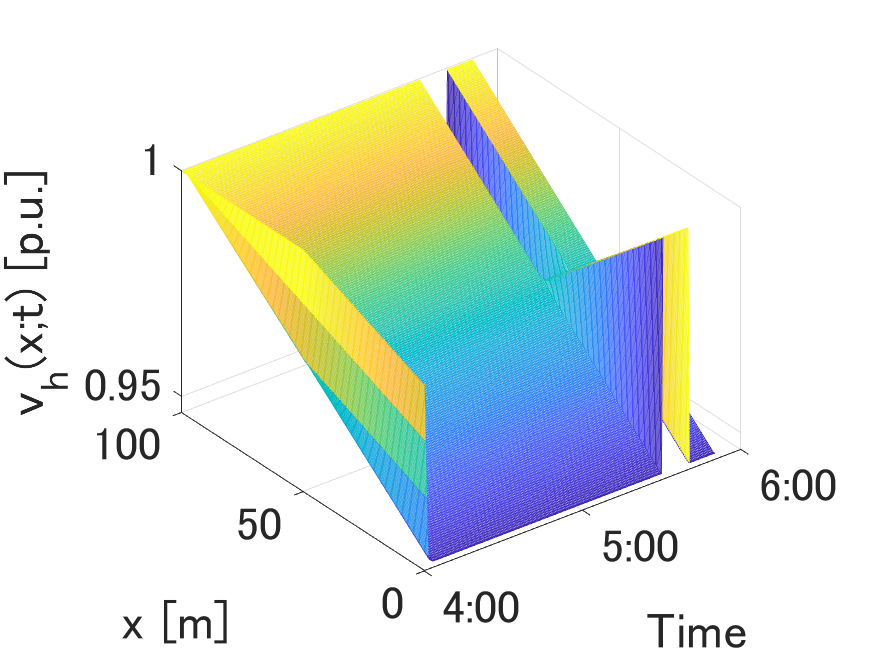}
}\\
\subfigure[Temperature of cable surface]{
\includegraphics[width=40mm]
{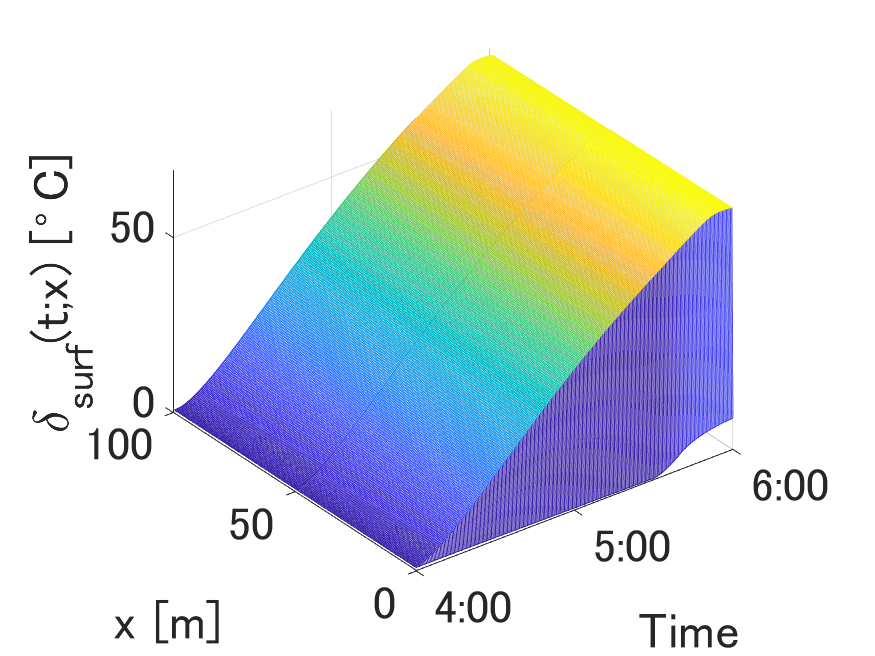}
}
\subfigure[Snow volume]{
\includegraphics[width=40mm]
{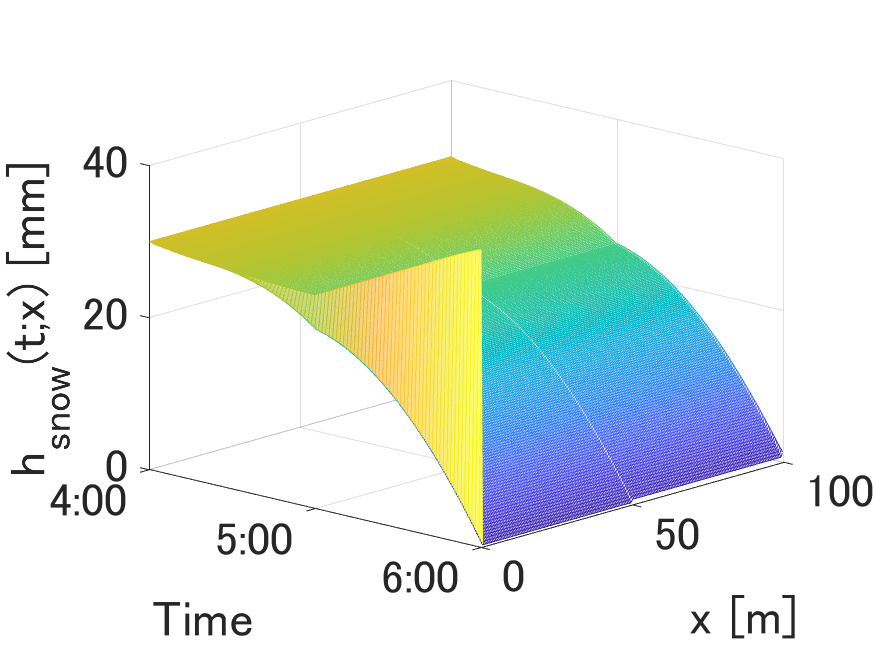}
}
\end{center}
\caption{
Spatial and temporal variation of the variables (Case~2)}
\label{simkekka2_mscs}
\end{figure}

\subsubsection*{(2) Temporal variation of switching and discharging}

We show the simulation results for the switches, the PV power generation equipment, and the battery storage in Fig.~\ref{simkekka2switch_mscs}. 
The temporal variation of the switching of the switch with subscript~$i$, i.e., {\tt On}, satisfying ${{\sigma_{i}(t)}}=1$ is shown in Fig.~\ref{simkekka2switch_mscs}~(a). From P.M.~4:00 to 5:30, 
Switch~2, which is powered by a battery storage, is set to {\tt On}. 
Since the owners of the heating cables and the battery storage are supposed to be the same, 
we can conclude that Switch~2 is chosen because it is less expensive than Switch~1, which is set to {\tt On}. 
From P.M.~5:30 to 5:40, Switch~1 is set to {\tt On}. 
{{The reason for selecting this switch is to promote the snow melting at the starting point~$x = 0$~[m].}} 
Then, from P.M.~5:50 to 6:00, the switch is set to {\tt Off} because the snow melting is almost completed. 

We show the temporal variation of the control input ~$\sigma_{{\rm pv}}(t)$ of the PV power generation equipment in Fig.~\ref{simkekka2switch_mscs}~(b). 
From P.M.~5:00 to 5:30, 
the equipment transmits power to the underground power distribution line. 
The reason for this is considered to be the increase in the power consumption of residential loads during this interval. 
Moreover, Fig.~\ref{simkekka2switch_mscs}~(c) shows the temporal variation of the subscript $i$ satisfying $\sigma_{{\rm battery},i}(t)=1$, i.e., the destination of the power from the battery storage. When Switch~2 is set to {\tt On}, the storage always transmits the power to the heating cable. 

The temporal variation of the amount of {{the active power}} stored in the battery storage is shown in Fig.~\ref{simkekka2switch_mscs}~(d). 
The initial amount of the {{active power}} in the battery storage is assumed to be $10$~[p.u.$\cdot$h]. From P.M.~4:00  to 4:30, the amount of the {{active power}} stored in the battery storage is increasing because the amount of the {{active power}} transmitted from the PV power generation equipment is larger than the amount of the electricity transmitted to the heating cable. 
However, from P.M.~4:30 p.m. to 5:30, the amount of the {{active power}} transmitted from the PV power generation equipment gradually decreases, and the amount of the {{active power}} in the battery storage also decreases.

\begin{figure}[htb]
\begin{center}
\subfigure[Temporal variation of switching control input]{
\includegraphics[width=40mm]
{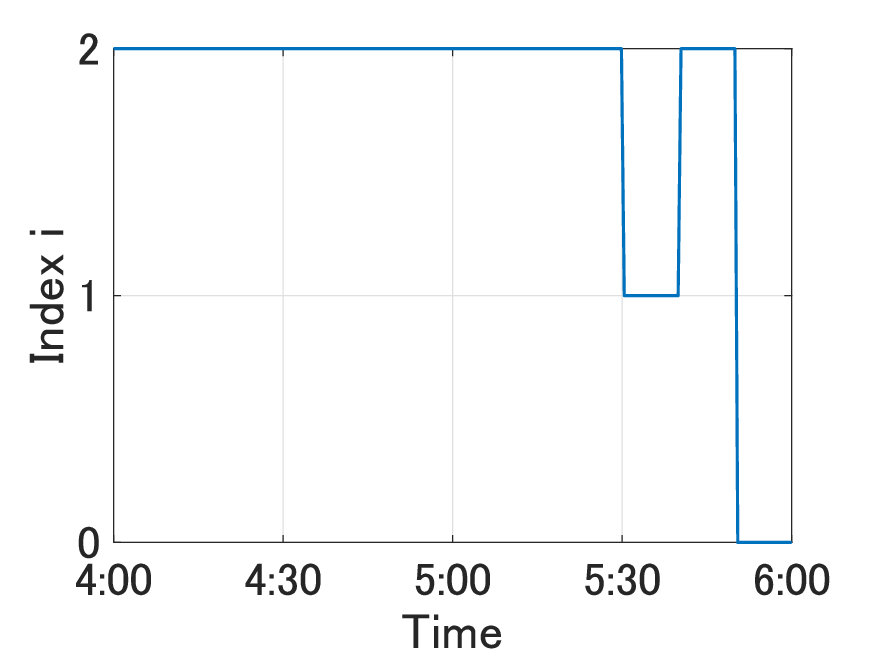}
}
\subfigure[Temporal variation of the destination of PV power]{
\includegraphics[width=40mm]
{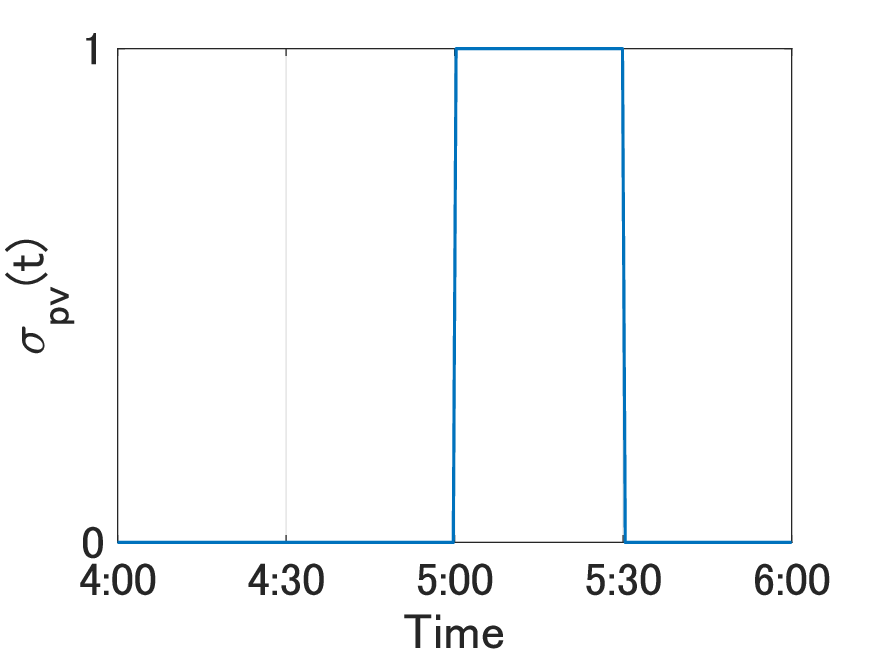}
}\\
\subfigure[Temporal variation of the destination of battery power]{
\includegraphics[width=40mm]
{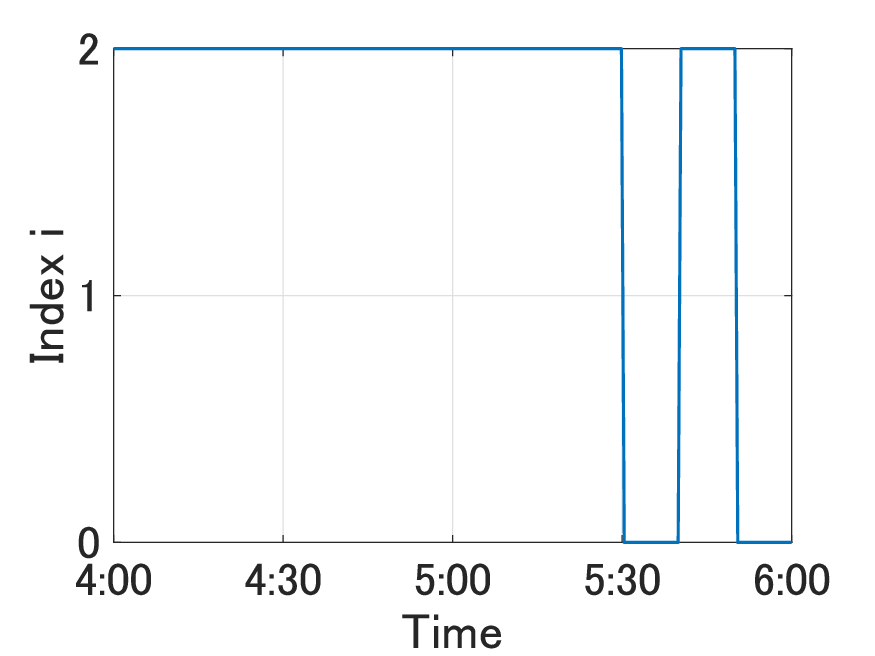}
}
\subfigure[Temporal variation of the amount of electricity in battery]{
\includegraphics[width=40mm]
{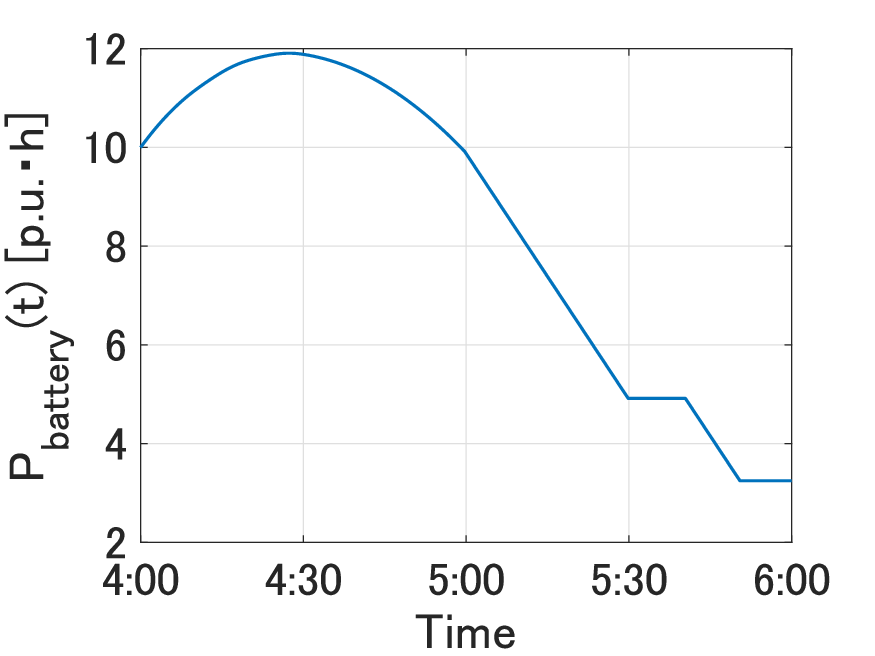}
}
\end{center}
\caption{
Temporal variation with respect to switches, PV, and battery storage~(Case~2)}
\label{simkekka2switch_mscs}
\end{figure}

\subsubsection*{(3) Comparison with and without battery storage}

In the following, we compare the costs and the voltage fluctuations with and without battery storage. 
Similarly to Case~1, the former and latter simulations correspond to the authors' previous study~\cite{sice} and this paper, respectively. 

Fig.~\ref{hikaku2}~(a) shows the switching when no battery storage is installed, 
and Fig.~\ref{hikaku2}~(b) shows the switching when a battery storage is installed. 
If no battery storage is installed, Switch~1 is selected between P.M.~4:00 and 5:40, 
and thus $16.7$~[p.u.$\cdot$h] of the {{active power}} is purchased. 
On the other hand, when the battery storage is installed, Switch~1 is selected from P.M.~5:30 to 5:40, 
and thus $1.7$~[p.u.$\cdot$h] of the {{active power}} is purchased. 
We can confirm that the introduction of the storage has significantly reduced the cost of using the heating cables. 

Figs.~\ref{hikaku2}~(c) and (d) show the voltage fluctuations without and with battery storage, respectively. From these figures, it can be confirmed that the voltage fluctuations are more suppressed when the battery storage is installed. 
This is because the battery storage optimizes the power flow according to the power consumption of the house and the amount of the PV power generation. 
{{Finally, Fig.~\ref{hikaku2} (e) shows the difference in the voltage fluctuations with and without battery storage. This figure shows that in Case~2, the installation of battery storage suppresses the voltage fluctuation through the considered period except for P.M.~4:50-5:00. 
During P.M.~4:50-5:00, there are large fluctuations in the effective power of the residential load, and the heating cable selects Switch~2 to give priority to snow melting. 
This is considered to have slightly affected the voltage fluctuations of the underground distribution line in the case with the battery storage. 
This observation well indicates the effectiveness of the proposed switching predictive control.}}

\begin{figure}[htb]
\begin{center}
\subfigure[Temporal variation of switching control input~(without battery storage)]{
\includegraphics[width=40mm]
{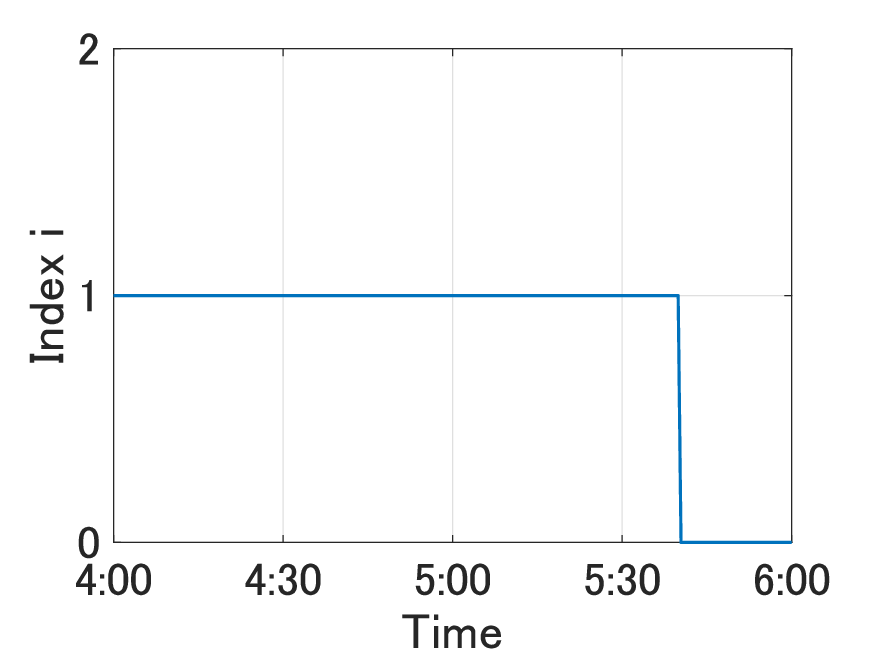}
}
\subfigure[Temporal variation of switching control input~(with battery storage)]{
\includegraphics[width=40mm]
{switch2_mscs.eps}
}\\
\subfigure[Voltage fluctuation~(without battery storage)]{
\includegraphics[width=40mm]
{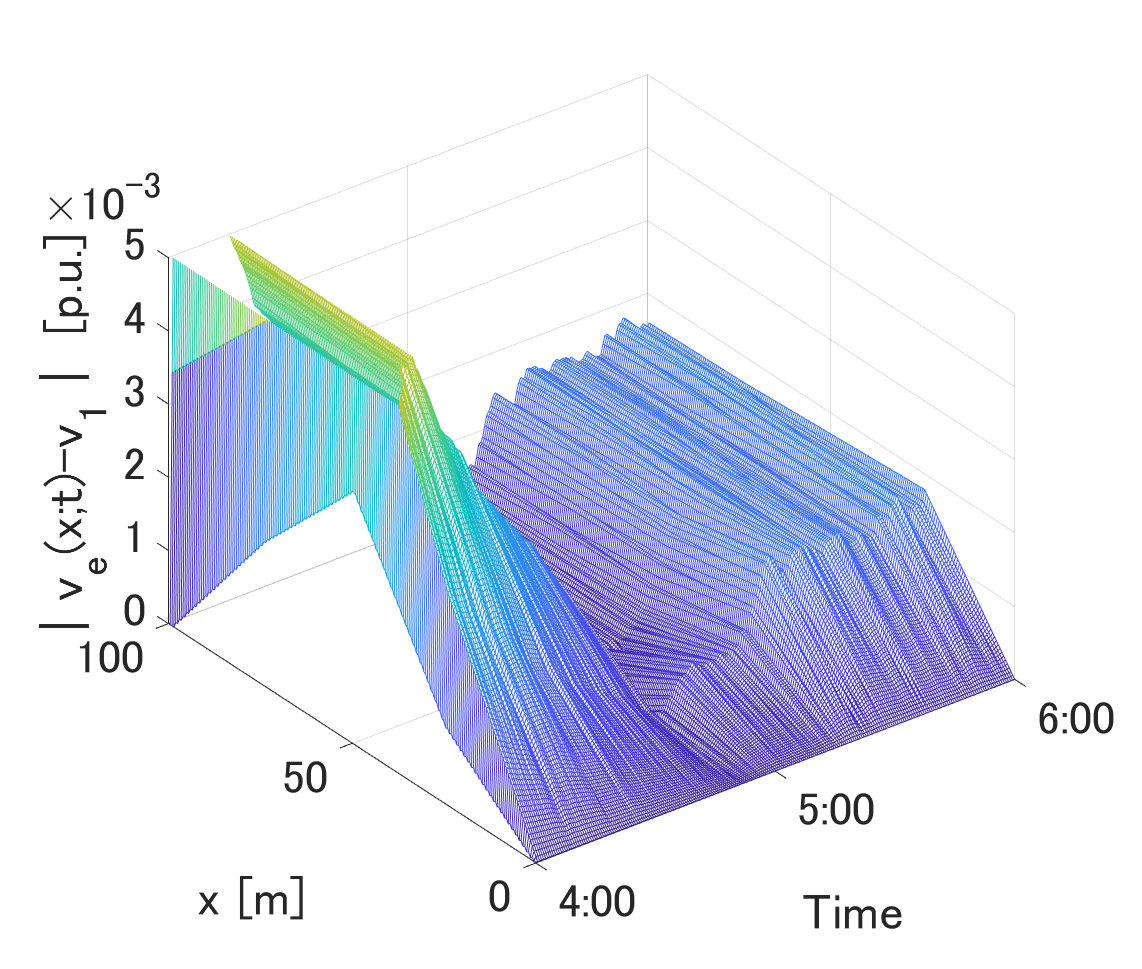}
}
\subfigure[{{Voltage fluctuation~(with battery storage)}}]{
\includegraphics[width=40mm]
{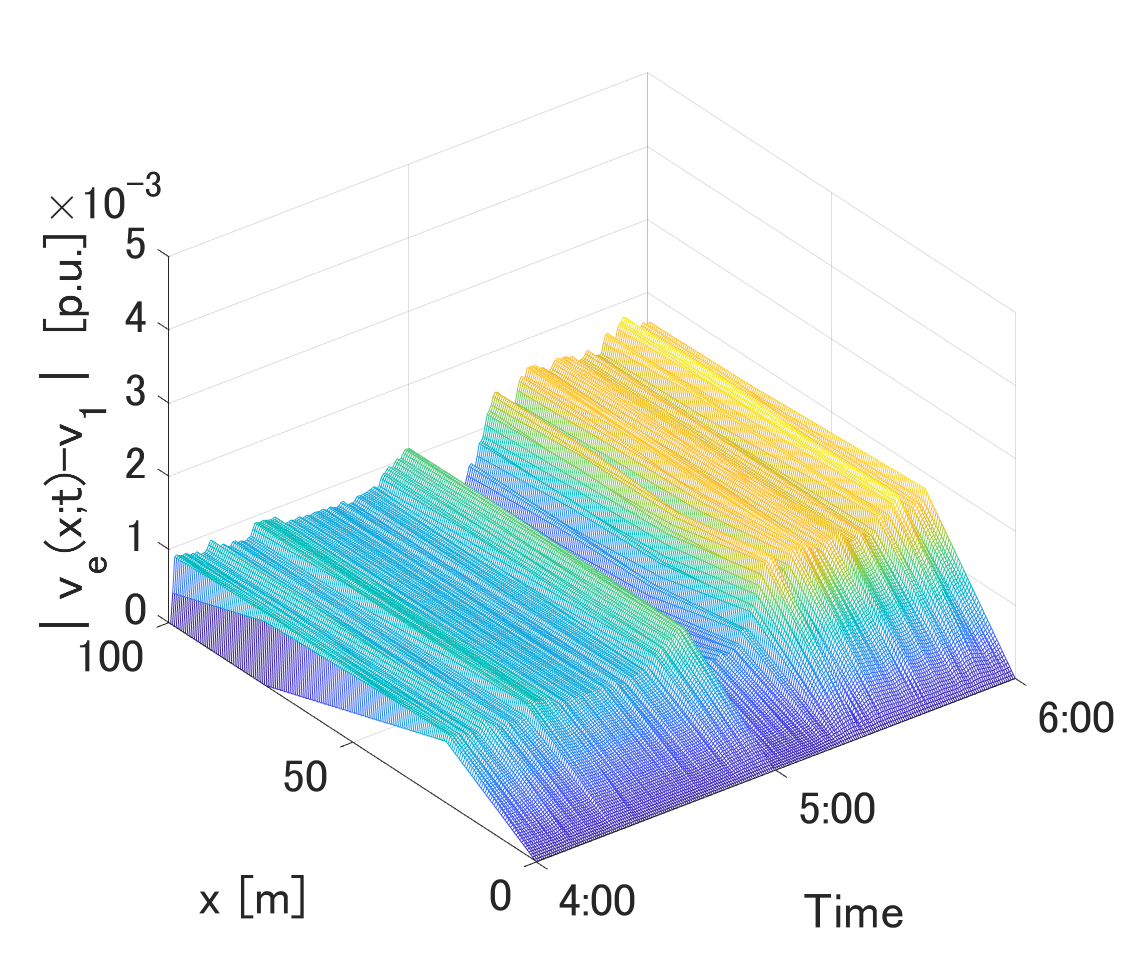}
}\\
\subfigure[{{Difference of voltage fluctuations}}]{
\includegraphics[width=40mm]
{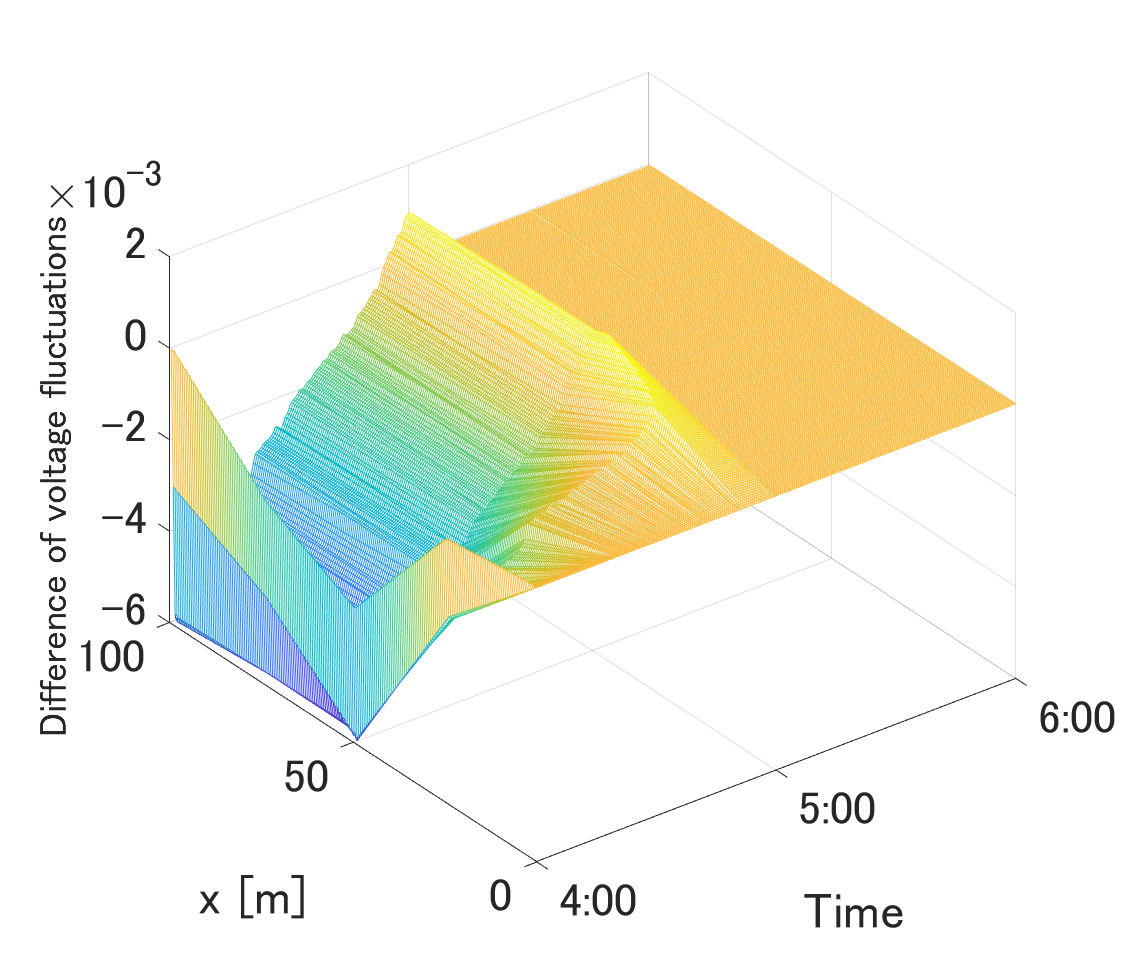}
}
\end{center}
\caption{
Comparison of switching control input and voltage fluctuation with respect to battery storage~(Case~2)}
\label{hikaku2}
\end{figure}

{{For the numerical simulations performed with the setting of Case~2, 
the computation times were 319.8~sec and 484.8~sec for Section~4.4 of the previous study~[9] and this paper, respectively.
Similarly to Case~1, although the computation time is slightly higher in this paper, 
the computation time required for a single switching is $484.8 \times \frac{10}{120} = 40.4$~[sec] for this paper. 
This is sufficiently less than the minimum switching time~$T_{\rm mini}=10$~[min] to complete the computation.
We can also see that the switching predictive control in this paper has no problem in real time implementation.}}

\section{Conclusions}
\label{sec:conc}

In this paper, we have studied the application of a battery storage to the predictive switching control of a power distribution system including road heating proposed by the authors~\cite{sice}. 
The predictive switching control including the switching of the PV power generation equipment and the battery storage has been proposed as a main result. 
We have demonstrated the effectiveness by numerical simulations using the actual time series data. 
As a conclusion, we have confirmed the efficiency of the introduction of battery storage compared to the authors' previous studies~\cite{sice}\cite{mscs}.

As a future work, 
{{the following issues (i)-(iv) can be considered.}}
\begin{itemize}
\item[{{(i)}}]
{{To}} improve the control performance of the proposed control, it is necessary to add constraints on the voltage and current amplitudes, 
and to consider appropriate weights for the evaluation functions. 
The maximum hands-off control~\cite{sper} can be introduced to reduce the number of switching events.
\item[(ii)]
{{It is desired}} to extend the proposed switching predictive control to the framework which also consider reactive power injection in order to suppress voltage phase delays and voltage fluctuations. 
A preliminary result has been obtained in the authors' study~\cite{sci}.
\item[{{(iii)}}]
{{In order to perform simulation verification that more accurately reflects reality for the switching predictive control proposed in this paper, 
it is necessary to derive a mathematical model of the power distribution system 
that explicitly incorporates the charging and discharging efficiency of the battery storage, 
as well as the power generation efficiency and conversion efficiency of PV power generation in relation to solar radiation. It is also desirable to explicitly include the installation cost of the battery storage in the evaluation functions considered in this paper. }}
\item[{{(iv)}}]
{{To improve the localized increase in snow volume observed in Subsections~\ref{sec:resu1} and \ref{sec:resu2}, 
the mathematical model of the considered power distribution system should explicitly incorporate phenomena such as horizontal thermal diffusion on the cable surface and normal thermal diffusion in the underground and snow volume. 
It is also necessary to extend the model to predictive switching control by explicitly incorporating the upper bound of the allowable snow volume in the constraints and evaluation functions.}}
\end{itemize}


\begin{thebibliography}{99}
\bibitem{road}
W.~J. Eugster: ``Road and bridge heating using geothermal energy-Overview and examples,''  
\textit{Proceedings European Geothermal Congress 2007}, 
2007.
\bibitem{stat}
W.B.~Yu, X.~Yi, M.~Guo, and L.~Chen: 
``State of the art and practice of pavement anti-icing and
de-icing techniques,'' 
\textit{Sciences in Cold and Arid Regions}, 
vol.~6, no.~1, pp.~14-21, 2014. 
\bibitem{eu01}
R.~Merrifield: ``Under-road heating system to keep Europe's highways ice-free,'' 
\textit{Horizon: The EU Research and Innovation Magazine, European Commission}, 
2019.
\bibitem{dist1}
W.H. Kersting: 
\textit{Distribution system modeling and analysis}, Fourth edition, Routledge, 2018.
\bibitem{powe1}
P.S. Kundur: \textit{Power system stability and control}, McGrow Hill, 1994.
\bibitem{nolta}
Y.~Muto, C.~Kojima, and Y.~Okura: ``Mathematical modeling of road heating system with underground power distribution line based on nonlinear ODE model,'' 
\textit{Nonlinear Theory and Its Applications}, IEICE, vol.~14, no.~2, pp.~378-402, 2023.
\bibitem{arxiv}
M.~Chertkov, S.~Backhaus, K.~Turtisyn, V.~Chernyak, and V. Lebedev: 
``Voltage collapse and ODE approach to power flows: Analysis of a feeder line with static disorder in consumption/production,'' \textit{arXiv}, 1106.5003, 2011.
\bibitem{haiden}
H.~Tadano, Y.~Susuki, and A.~Ishigame: 
``Asymptotic assessment of distribution voltage profile using a nonlinear ODE model,'' \textit{Nonlinear Theory and Its Applications}, IEICE, 
vol.~13, no.~1, pp.~149-168, 2022.
\bibitem{sice}
Y.~Muto, C.~Kojima, and Y.~Susuki: 
``Switching Predictive Control of Power Distribution System Including Road Heating,'' 
\textit{Transactions of the Society of Instrument and Control Engineers}, 
vol.~60, no.~6, pp.~384-396, 2024 (in Japanese).
\bibitem{esaim}
F.~Bernardin and A.~Munch: Modeling and optimizing a
road de-icing device by a nonlinear heating, 
\textit{ESAIM: Mathematical Modelling and Numerical Analysis}, vol.~53, no.~3, pp.~775-803 (2019)
\bibitem{mscs}
Y.~Muto, H.~Akutsu, C.~Kojima, and Y. Susuki: 
``Application of Battery Storage to Switching Predictive Control of Power Distribution Systems Including Road Heating,'' 
\textit{Proceedings of the 11th SICE Multi-symposium on Control Systems}, 3A2-3, 2024  (in Japanese).
\bibitem{sug:eval1}
H.~Sugihara, T.~Funaki, and N.~Yamaguchi: 
``Evaluation method for real-time dynamic line ratings based on line current variation model for representing forecast error of intermittent renewable generation,'' 
\textit{Energies}, 
vol.~10, no.~4, 503, 2017.
\bibitem{ino:effe1}
A.~Inomata, R.~Hara, T.~Oyama, and I.~Kurihara: 
``Effectiveness of application of dynamic rating to transmission system,'' 
\textit{IEEJ Transactions on Power and Energy}, 
vol.~126, no.~1, pp.~36-42, 2006 (in Japanese).
\bibitem{soudenrosu}
C.~Kojima, Y.~Muto, and Y.~Susuki: 
``On dissipativity of nonlinear ODE model of distribution voltage profile,'' 
\textit{Proceedings of IFAC World Congress 2023}, 
pp.~7031-7036, 
2023.
\bibitem{itijigenkyokai}
I.~Gushiken and N.~Tosaka: 
``Boundary element analysis of 1-D unsteady thermal convection problem,Transactions of Boundary Element Method,'' 
\textit{Japan Society for Computational Methods in Engineering}, 
vol.~19, no.~017\_021111, 2002 (in Japanese).
\bibitem{netusyusi}
H.~Takimoto, A.~Ogura, M.~Yoshida, K.~Takase, and T.~Maruyama: 
``Analysis of snowpack accumulation and melting with a heat balance approach - Role of heat flux from underground in melting snow,'' 
\textit{Transactions of The Japanese Society of Irrigation, Drainage and Rural Engineering}, 
vol.~82, no.~4, pp.~191-200, 2014 (in Japanese).
\bibitem{odesim}
Y.~Susuki, S.~Baek, Y.~Ota, and T.~Hikihara: 
``Computer Simulation of Distribution Voltage Profile Using a Nonlinear ODE,'' \textit{IEICE Technical Report}, 
vol.~116, no.~215, NLP2016-47, pp.~15-20, 2016 (in Japanese).
\bibitem{phouse}
Nagoya University, 
\textit{Nagoya University Open Data for EMS Evaluation}, \url{http://data.v2x-ems.nagoya/down/sample}, 
2023  (in Japanese). 
\bibitem{pvdata}
Hokuriku Electric Power Transmission and Distribution Company: 
\textit{Hokuriku Area Electricity Forecast}, \url{https://www.rikuden.co.jp/nw/denki-yoho/}, 2022 (in Japanese). 
\bibitem{sundata}
NEDO: \textit{Database for Solar Radiation}, \url{https://appww2.infoc.nedo.go.jp/appww/index.html}, 
2017 (in Japanese). 
\bibitem{temp}
Japan Meteorological Agency: \textit{Historical Data Search}, \url{https://www.data.jma.go.jp/obd/stats/etrn/index.php}, 
2023 (in Japanese). 
\bibitem{sper}
M.~Nagahara, D.E.~Quevedo, and D.~Ne$\check{\rm s}$i$\acute{\rm c}$, ``Maximum hands-off control: A paradigm of control effort minimization, \textit{IEEE Transactions on Automatic Control}, vol.~61, no.~3, pp.~735-747 (2016)
\bibitem{sci}
R.~Shima, Y.~Muto, H.~Akutsu, C.~Kojima, and Y.~Susuki: 
``Switching Predictive Control of Power Distribution System Including Road Heating with Reactive Power Injection'', 
\textit{Proceedings of the 69th Annual Conference of ISCIE}, {{pp.~946-950}}, 2025  {{(in Japanese)}}.
\end{thebibliography}
\end{document}